\documentclass[twocolumn,twocolumn]{IEEEtran}

\usepackage[T1]{fontenc}
\usepackage{color}
\usepackage{array}
\usepackage{float}
\usepackage{mathrsfs}
\usepackage{amsmath}
\usepackage{amssymb}
\usepackage{graphicx}

\usepackage{fancyhdr}

\usepackage[unicode=true,
 bookmarks=true,bookmarksnumbered=true,bookmarksopen=true,bookmarksopenlevel=1,
 breaklinks=false,pdfborder={0 0 0},pdfborderstyle={},backref=false,colorlinks=false]
 {hyperref}
\hypersetup{pdftitle={Your Title},
 pdfauthor={Your Name},
 pdfpagelayout=OneColumn, pdfnewwindow=true, pdfstartview=XYZ, plainpages=false}

\makeatletter

\floatstyle{ruled}
\newfloat{algorithm}{tbp}{loa}
\providecommand{\algorithmname}{Algorithm}
\floatname{algorithm}{\protect\algorithmname}

\usepackage[caption=false,font=footnotesize]{subfig}
\usepackage{algorithm}
\usepackage{algorithmic}

\usepackage{multirow} 
\usepackage{amsmath} 
\usepackage{xcolor}

\allowdisplaybreaks[4]

\ifCLASSOPTIONcompsoc
\usepackage[caption=false,font=normalsize,labelfont=sf,textfont=sf]{subfig}
\else
\usepackage[caption=false,font=footnotesize]{subfig}
\fi

\usepackage{booktabs}

\usepackage{cite}
\usepackage{bm}
\usepackage{algorithmic}
\usepackage{algorithm}
\usepackage{graphicx}
\IEEEoverridecommandlockouts

\usepackage{lettrine}

\usepackage{geometry}
\@ifundefined{showcaptionsetup}{}{%
 \PassOptionsToPackage{caption=false}{subfig}}
\usepackage{subfig}
\makeatother

\usepackage{cases}

\begin{document}
\hypersetup{hidelinks}

\title{\textcolor{black}{NFV-Enabled Service Recovery in Space-Air-Ground Integrated Networks: A Matching Game Based Approach}}

\author{\IEEEauthorblockN{Ziye Jia, \IEEEmembership{Member, IEEE,} Yilu Cao, Lijun He, \IEEEmembership{Member, IEEE, }Guangxia Li, Fuhui Zhou, \IEEEmembership{Senior Member, IEEE, }\\ Qihui Wu, \IEEEmembership{Fellow, IEEE, }and Zhu Han, \IEEEmembership{Fellow, IEEE}}
\thanks{Ziye Jia is with the College of Electronic and Information Engineering, Nanjing University of Aeronautics and Astronautics, Nanjing 211106, China, and also with the State Key Laboratory of ISN, Xidian University, Xian 710071, China (e-mail: jiaziye@nuaa.edu.cn).}
\thanks{Yilu Cao, Guangxia Li, Fuhui Zhou and Qihui Wu are with the College of Electronic and Information Engineering, Nanjing University of Aeronautics and Astronautics, Nanjing 211106, China (e-mail: caoyilu@nuaa.edu.cn; guangxia@nuaa.edu.cn; zhoufuhui@nuaa.edu.cn; wuqihui@nuaa.edu.cn).}
\thanks{Lijun He is with the School of Information and Control Engineering, China University of Mining and Technology, Xuzhou 221116, China (e-mail: lijunhe@cumt.edu.cn).}
\thanks{Zhu Han is with the Department of Electrical and Computer Engineering at the University of Houston, Houston, TX 77004 USA, and also with the Department of Computer Science and Engineering, Kyung Hee University, Seoul, South Korea, 446-701 (e-mail: hanzhu22@gmail.com).}}


\maketitle

\fancyhf{}

\fancyhead[R]{\scriptsize\thepage}

\renewcommand{\headrulewidth}{0pt}

\pagestyle{fancy}

\thispagestyle{fancy}
\begin{abstract}
    To achieve ubiquitous connectivity of the sixth generation communication, the space-air-ground integrated network (SAGIN) is a popular topic. However, the dynamic nodes in SAGIN such as satellites and unmanned aerial vehicles, may be fragile and out of operation, which can potentially cause service failure. Therefore, the research on service recovery in SAGIN under situations of resource failure is critical. In order to facilitate the flexible resource utilization of SAGIN, the network function virtualization technology (NFV) is proposed to be employed. Firstly, the task management is transformed into the deployment of service function chains (SFCs). Then, we design an NFV-based SFC recovery model in SAGIN in the face of resource failure, so that tasks can quickly select alternative resources to complete deployments. Moreover, the problem of SFC recovery is formulated to minimize the total time consumption for all completed SFCs. Since it is an NP-hard integer linear programming problem, we propose the efficient recovery algorithm based on the matching game. Finally, via various simulations, the effectiveness of the proposed algorithm and its advantages are verified, where the total time consumption is optimized by about 25\%, compared with other benchmark methods.
    
\end{abstract}
\begin{IEEEkeywords}
    Space-air-ground integrated network, network function virtualization, service function chain, resource failure, service recovery, matching game.
\end{IEEEkeywords}

\newcommand{\CLASSINPUTtoptextmargin}{0.8in}

\newcommand{\CLASSINPUTbottomtextmargin}{1in}

\section{Introduction}

\lettrine[lines=2]{T}{HE} space-air-ground integrated network (SAGIN) is a popular research issue in recent years, which is one key component of the sixth generation communication technology \cite{8368236}, \cite{cheng20226g}. SAGIN is mainly composed of satellites, unmanned aerial vehicles (UAVs), high-altitude platforms, and ground stations \cite{9722775}. SAGIN can provide a large amount of information services in environmental monitoring, traffic management, industry and agriculture, and disaster rescue \cite{10579820}.

However, as a multi-layer and highly heterogeneous network, SAGIN has diverse resource types and complex resource structures, such as the high dynamic periodic movement of satellite nodes, the flexibility and controllability of UAV nodes, and the different resource capabilities of multiple network nodes \cite{8612450}. In addition, the resources of different nodes are limited to complete certain specific tasks \cite{10572013}. Hence, diverse networks cannot share resources, resulting in poor task processing ability and low resource utilization \cite{jia2021vnf}, \cite{10440193}. Moreover, the issues involving network malfunction and resource failure, affect the execution and transmission of tasks in SAGIN \cite{9628162}. These problems lead to that tasks can not be transmitted or processed smoothly. As a result, the completion time of tasks increases, which affects the network efficiency. Therefore, effective and robust task recovery is necessary to deal with the resource failure \cite{Liu2023}.

Accordingly, the network function virtualization (NFV) technology is introduced into SAGIN, which implements by decoupling the software and hardware devices to shield the technical differences in various devices, so as to realize the interconnection and resource sharing of SAGIN \cite{herrera2016resource}, \cite{10089243}. The entire NFV architecture consists of network functions virtualization infrastructure (NFVI),  management and orchestration, and virtual network functions (VNFs) \cite{10061634}. Among them, VNFs are the software application that realizes the network function, and implemented on NFVI \cite{10587207}, \cite{10090468}. A task that contains multiple VNFs connecting sequentially can be represented as a service function chain (SFC) \cite{10638237}, \cite{20233814771268}. Therefore, the SFC deployment provides a new mechanism for the elastic resource management of SAGIN, but it also brings the following new challenges.

\begin{itemize}
    \item[$\bullet$] How to achieve a unified representation of resources across different domains and time periods is a key challenge, which hinders the ability to create an effective deployment model for SFCs. Moreover, the heterogeneity of resources across various domains and the temporal variations aggravate the difficulties to standardize the resource representation, which is essential for the deployment of SFCs.
    \item[$\bullet$] The highly dynamic networks are susceptible to network malfunctions and resource failures. These instabilities may disrupt the deployment of SFCs, leading to service interruptions and degraded performance. It is essential to manage and mitigate these risks to maintain the SFC deployments in such volatile environments.
    \item[$\bullet$] It is challenging to design the robust recovery mechanism to handle resource failures and ensure the redeployment of SFCs. In the event of resource failure, it is crucial to implement an effective strategy for quickly recovering to minimize time consumption and maintain service continuity. Besides, the resource reallocation and efficient redeployment strategies should also be considered.
\end{itemize}

To overcome the above challenges, in this paper, firstly, we model the SAGIN network and propose a reconfigurable time expansion graph (RTEG) that divides the time horizon into several small time slots. When the time slots are short enough, each node is regarded as quasi-static in a time slot. Therefore, the same node can be considered as different nodes in adjacent time slots, so as to facilitate the analysis of models and problems. Then, we convert tasks as SFCs with multiple VNFs, and propose a model of SFC deployment in SAGIN for the transmission and processing of tasks. Due to the dynamic nature and instability of the network, the node failure may occur in each time slot. At the same time, the associated links are disconnected, and the SFCs can not continue to transmit on the original links. Then, the problem of service recovery in the case of resource failure is established to minimize the total time consumed to complete all SFC deployments. Since the formulated problem is in the form of integer linear programming (ILP), it is NP-hard and intractable to be solved directly \cite{vazirani2001approximation}. Hence, we propose the algorithm based on matching game.

In summary, the contributions of this paper are listed as follows:
\begin{itemize}
    \item[$\bullet$] We present the SFC deployment model based on RTEG. The time period is divided into several small time slots, to handle the high dynamic of SAGIN. Then, it is transformed into a quasi-static network in each time slot, which is convenient for analyzing the model of SFC deployments.
    \item[$\bullet$] We focus on the service recovery confronted with resource failure in SAGIN, including the failure of UAV and satellite nodes, as well as node-related link failures. A detailed SFC deploy and recovery model, including strategies for dealing with resource failures is proposed.
    \item[$\bullet$] In order to solve the formulated problem, we propose an algorithm based on matching game to deal with unpredictable resource failures, so that SFCs can quickly handle emergencies caused by resource failures, and be transmitted and deployed smoothly. 
    \item[$\bullet$] Simulation results show the effectiveness and advantages of the proposed algorithm, where the total time consumption is optimized by about 25\%, compared with other benchmark algorithms.
\end{itemize}

The structure of this paper is organized as follows. Section \ref{sec:Related-Work} discusses the related work. In Section \ref{sec:System-Model}, we introduce the system model, followed by the problem formulation in Section \ref{sec:Problem-Formulation}. The algorithm designs are detailed in Section \ref{sec: DRL algorithm}. Section \ref{sec:Simulation-Results} presents the simulation results along with the analysis. Finally, conclusions are drawn in Section \ref{sec:Simulation-Results}.

\section{Related Work\label{sec:Related-Work}}



There exist several studies exploring SFC or VNF deployment in SAGIN. For example, the authors in \cite{9951143} applied federated learning algorithms to address the SFC embedding challenge in SAGIN and adjusted SFC configurations to minimize service blocking rates. The authors in \cite{AKYILDIZ2019134} introduced a cyber-physical system that integrates ground, air, and space layers using software-defined networking (SDN) and NFV techniques. In \cite{fi16010027}, the authors proposed a hierarchical resource management structure for SAGIN, which integrated SDN, NFV, and multi-access edge computing to manage heterogeneous network resources. In \cite{9062531}, the authors proposed a heuristic greedy algorithm to tackle the SFC planning issue in a reconfigurable service provisioning framework for SAGIN. The work of \cite{10398221} proposed a SAGIN architecture with edge intelligence to improve the capabilities of communication, computing, sensing, and storage. It introduced a new deep reinforcement learning-based algorithm for resource allocation and computation offloading. In \cite{9749937}, the authors developed a service model by segmenting network slices and suggested an SFC mapping approach based on delay prediction. The authors in \cite{9351537} focused on dynamic VNF mapping and scheduling within SAGIN, and proposed two Tabu search-based algorithms for achieving near-optimal solutions. In \cite{10123085}, the authors proposed a dynamic network architecture for integrating terrestrial and non-terrestrial networks, addressing challenges of satellite mobility and communication delays. It optimized the allocation of VNFs among LEO CubeSats using three kinds of heuristic algorithms, achieving the near-optimal performance in simulations. These papers considered various models and algorithms in SAGIN to solve the deployment problems of SFCs or VNFs. However, the resource failures are not well considered, which is an important problem in dynamic SAGIN. In the dynamic network environment, nodes and links are easy to fail due to the environmental uncertainty, resulting in the tasks on nodes and links can no longer be effectively transmitted and processed, respectively.

\begin{table*}[!t]
    \renewcommand\arraystretch{1.3}
	\begin{center}
		\caption{EVALUATION OF THE RELATED WORK} \label{evaluation of the related work}
        \begin{tabular}{|c|c|c|c|c|c|}
            \hline
            \multirow{2}{*}{References} & \multicolumn{5}{c|}{Requirements} \\ \cline{2-6} & Deployment & Scheduling & Resource failure & Terrestrial networks & Non-terrestrial networks\\
            \hline
            \hline
            \cite{9951143}, \cite{9351537} & \checkmark & \checkmark & - & - & \checkmark \\
            \hline
            \cite{AKYILDIZ2019134}, \cite{fi16010027}, \cite{10398221}, \cite{9749937}, \cite{10123085} & \checkmark & - & - & - & \checkmark \\
            \hline
            \cite{9062531} & - & \checkmark & - & - & \checkmark \\
            \hline
            \cite{8463632}, \cite{9585385}, \cite{10179962}, \cite{8923413}, \cite{9296232}, \cite{10533654} & - & - & \checkmark & \checkmark & - \\
            \hline
            \cite{7987282}, \cite{7898396} & \checkmark & - & \checkmark & \checkmark & -\\
            \hline
            Our work & \checkmark & \checkmark & \checkmark & - & \checkmark \\
            \hline
		\end{tabular}
	\end{center}
\end{table*}

\begin{table}[!t]
    \renewcommand\arraystretch{1.3}
	\begin{center}
		\caption{KEY NOTATIONS} \label{key notations}
		\begin{tabular}{|p{2cm}|p{6cm}|}
			\hline
			Symbol& Description \\
			\hline
            \hline
			$\mathcal{G} =(\mathcal{E},\mathcal{Y})$ & SAGIN graph composed of nodes $\mathcal{E}$ and links $\mathcal{Y}$.\\
            \hline
			$T$, $t$, $\mathscr{T}$, $\tau$ & Set of time slots, order number of time slots, total number of time slots, and time slot length.  \\
            \hline
            $\mathcal{V}_k$, $\mathcal{K}$, $l_k$ & SFC of the $k$-th task, the total number of tasks, and the total number of VNFs in the $k$-th SFC. \\
            \hline
            $i^t$, $v_k^m$ & The node $i$ in time slot $t$, and the $m$-th VNF of SFC $\mathcal{V}_k$. \\
			\hline
            $x_{v^m_k, i^t} $ & Binary variable indicating whether VNF $v^m_k$ of SFC $\mathcal{V}_k$ is deployed on node $i^t$.\\
            \hline
            $y^k_{(i^t,j^t)}$ & Binary variable indicating whether SFC $\mathcal{V}_k$ is deployed on link $(i^t_i,j^t)$.  \\
            \hline
            $z_{(i^t,i^{t+1})}^k $ & Binary variable indicating whether SFC $\mathcal{V}_k$ is stored on $i^t$ from $t$ to $t+1$.\\
            \hline
            $w^m_k$ & Binary variable indicating whether VNF $v^m_k$ of SFC $\mathcal{V}_k$ needs to be redeployed.\\
            \hline
            $\varphi _{i^t}$ & Computation ability of node $i^t$.\\
            \hline
            $\Delta_k$ & Data amount of SFC $\mathcal{V}_k$.\\
            \hline
            $\sigma _{v_k^m}$ & Computing resource consumed by VNF $v_k^m$.\\
            \hline
            $e^c_{i^t}$ & Energy consumption per unit of computing resource on node $i^t$.\\
            \hline
		\end{tabular}
	\end{center}
\end{table}

The resource failure and solutions based on NFV in terrestrial networks have been studied. For example, the authors in \cite{8463632} presented a framework for provisioning SFC request availability in multi-layered, heterogeneous-failure environment of a data center, and optimize resource usage. The authors in \cite{9585385} presented a dynamic virtual resource allocation mechanism for NFV-enabled vehicular and the fifth generation mobile communication technology (5G) networks. The mechanism was designed to maintain service performance despite network element failures. In \cite{10179962}, the authors introduced a digital twin-based scheme for SFC failure localization, involving failure classification and root cause analysis. In \cite{8923413}, the authors investigated quick VNF recovery using diversity coding in 5G networks, which avoided retransmissions and reduce capacity costs. It improved the reliability and reduced delays. The authors in \cite{9296232} tackled the service chain composition in NFV systems, considering user competition and resource failures. In \cite{10533654}, the authors introduced the VNF restoration problem in NFV and proposed the online recovery algorithm, which aimed to maximize the total weight of recovered services, prioritizing the restoration of the most important services during failures. In \cite{7987282}, the authors introduced a decision tree-based algorithm for VNF placement and chaining to mitigate penalties from link failures by Monte-Carlo Tree Search. The authors in \cite{7898396} presented a recovery approach for virtual networks affected by substrate node failures. They offered fair and priority-based recovery methods, and proposed a heuristic algorithm to solve the problem. However, the above papers only studied the fixed and stable physical terrestrial network, without considering the dynamic and heterogeneous nature of SAGIN.

As analyzed above, the problems of SFC deployment in SAGIN, and resource failure and recovery faced by NFV and SFCs in terrestrial networks are well studied. However, as far as the authors' knowledge, the research on recovery of SFCs in SAGIN is not comprehensive. Therefore, in this paper, we consider the dynamic resource instability of multi-layer SAGIN, propose the corresponding SFC deployment and recovery model in the case of resource failure, and design the algorithm based on matching game. The comparisons between our work and the related works are shown in Table \ref{evaluation of the related work}.

\begin{figure}[!t]
    \centerline{\includegraphics[width=9.9cm]{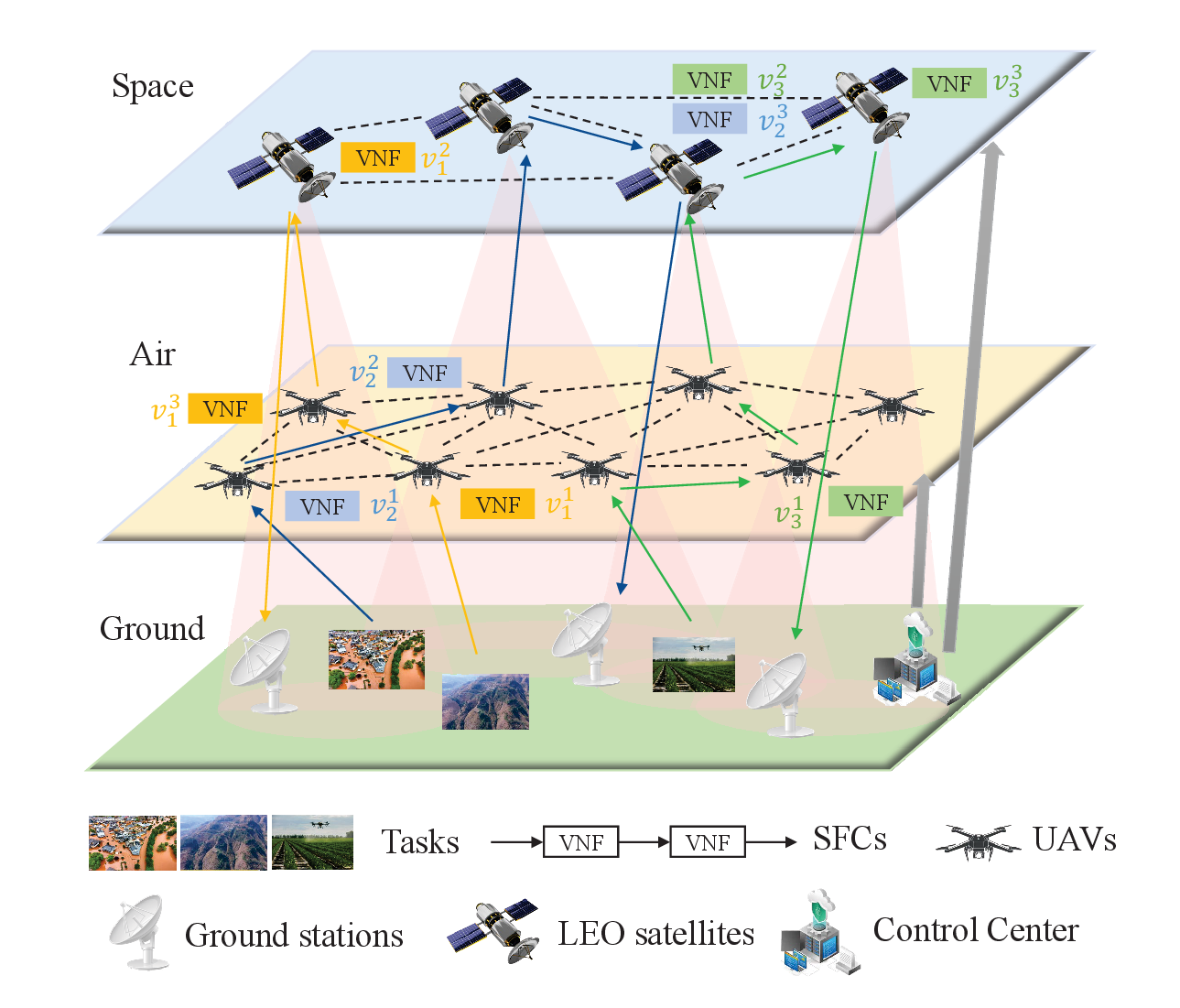}}
   \caption{Scenario of SFC deployments in SAGIN.}\label{fig1}
\end{figure}

\section{System Model\label{sec:System-Model}}
\begin{figure*}[!t]
    \centerline{\includegraphics[width=18cm]{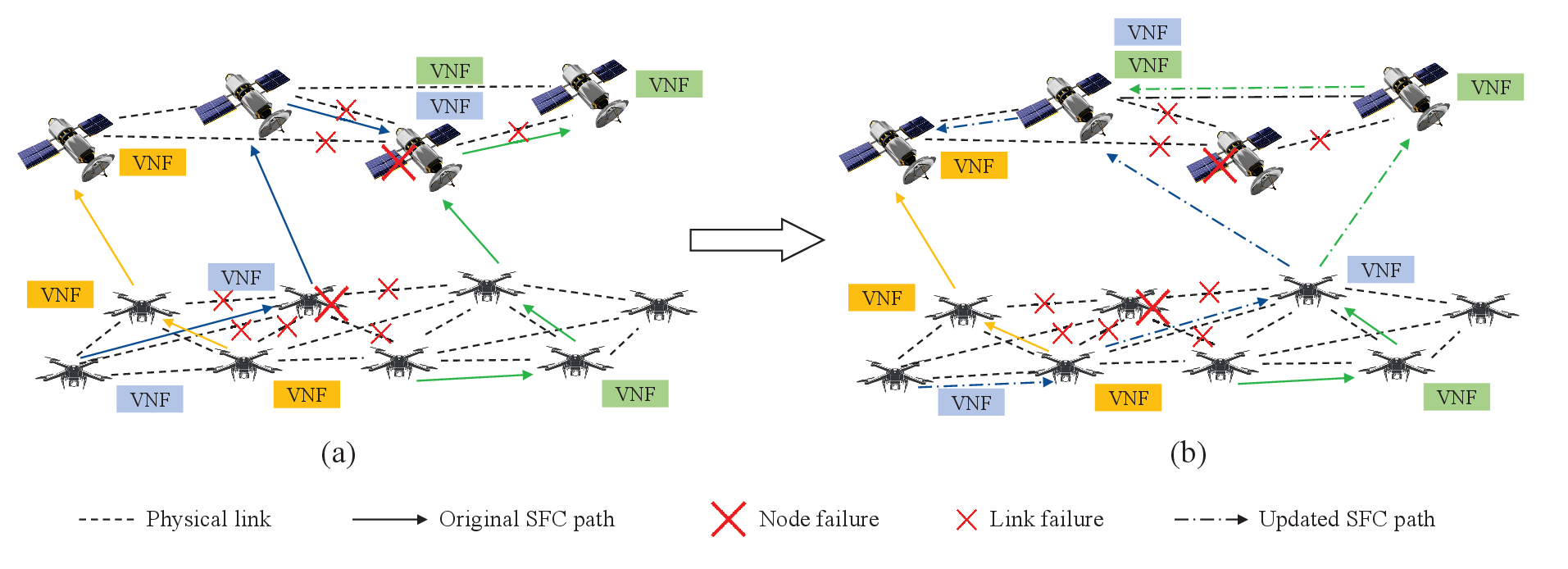}}
   \caption{SFC redeployment with failing nodes and links. (a) SFC deployments in SAGIN before resource failure. (b) SFC re-deployments corresponding to (a).\label{fig2}}
\end{figure*}

\begin{figure}[!t]
	\centering
	{\includegraphics[width=.9\columnwidth]{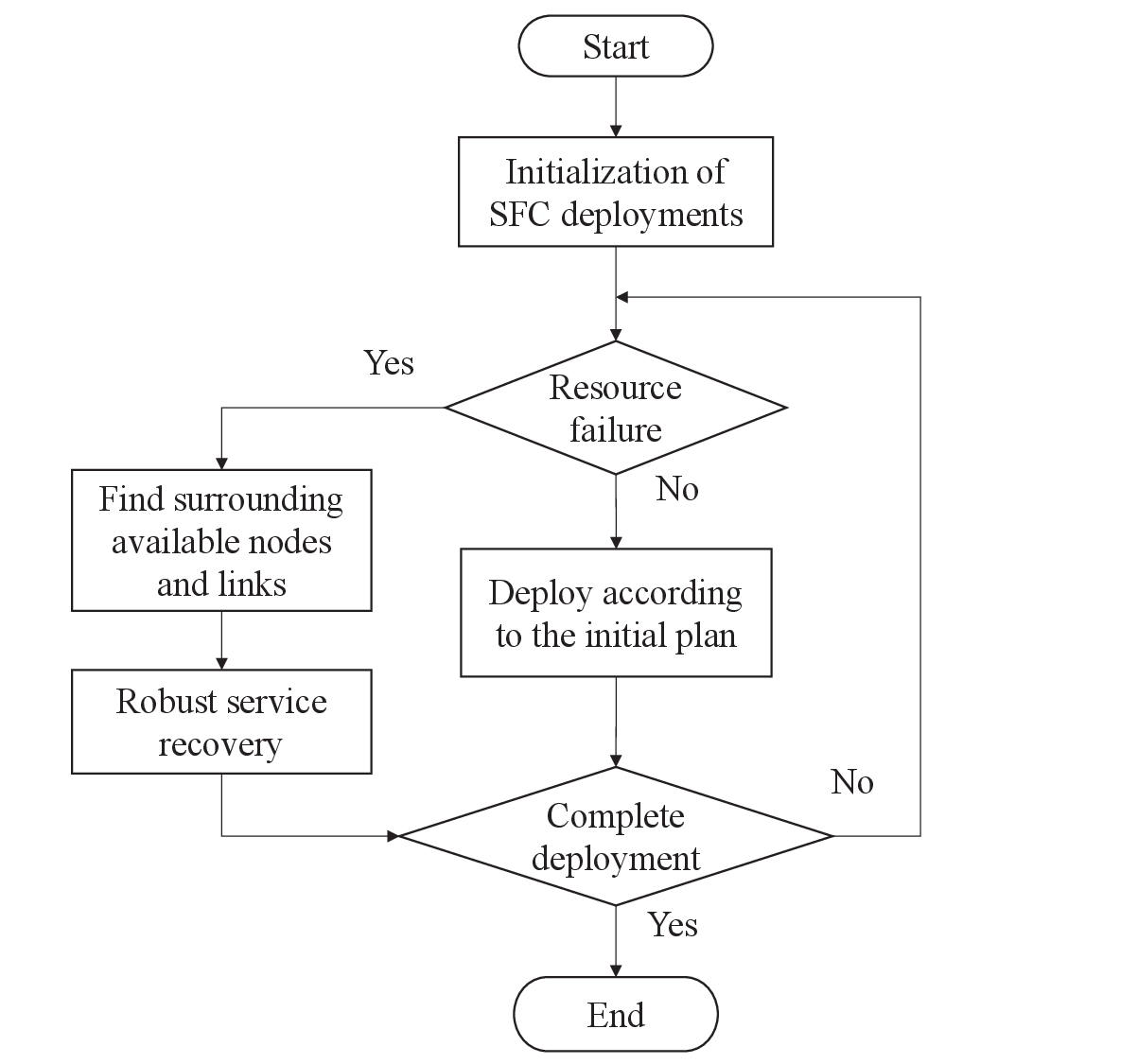}}
	\caption{The flow diagram of SFC redeployment.}
    \label{fig:flow diagram}
\end{figure}

In this section, the network model, channel model, and
energy model are respectively elaborated. Key notations used in this paper are listed in Table \ref{key notations}.

\subsection{Network Model}

As shown in Fig. \ref{fig1}, the considered SAGIN scenario is divided into three layers, containing low earth orbit (LEO) satellites, UAVs, and ground stations. Firstly, we propose RTEG and divide the time period into $T$ small time slots $t$, $t \in T$, where $\tau$ represents the length of each time slot. When the time slot is small enough, it can be assumed that UAV nodes and satellite nodes are quasi-static in time slot $t$. Then, we set the entire network as $\mathcal{G}=(\mathcal{E},\mathcal{Y})$, where $\mathcal{E}=\mathcal{E}_g \cup \mathcal{E}_u \cup \mathcal{E}_s$ represent nodes of ground users, UAVs, and satellites, respectively. $i^t \in \mathcal{E}$ represents the $i$-th node in time slot $t$. The link $\mathcal{Y}=\mathcal{Y}_l \cup \mathcal{Y}_t$, where $\mathcal{Y}_l=\mathcal{Y}_{gu} \cup \mathcal{Y}_{uu} \cup \mathcal{Y}_{us} \cup \mathcal{Y}_{ss} \cup \mathcal{Y}_{sg} \cup \mathcal{Y}_{ug}$ denotes links between two nodes, which include ground-to-UAV (G2U), UAV-to-UAV (U2U), UAV-to-satellite (U2S), satellite-to-satellite (S2S), satellite-to-ground (S2G), and UAV-to-ground (U2G) links, respectively, $(i^t,j^{t}) \in \mathcal{Y}_l$. Then, we set link $\mathcal{Y}_t=\{(i^t,i^{t+1})|i^t \in \mathcal{E}, t \in T\}$ to represent the storage link of SFCs on the same node $i$ from $t$ to its adjacent time slot $t+1$. 

In addition, the SFC is set as $\mathcal{V}_k =\{ v^1_k \rightarrow v^2_k \rightarrow ... \rightarrow v^m_k \rightarrow ... \rightarrow v^{l_k}_k\}$, where $k$ represents the order number of SFCs, $k \in \mathcal{K}$, and $\mathcal{K}$ is the total number of SFCs. $v^m_k$ denotes the $m$-th VNF in the $k$-th SFC, and $l_k$ indicates the total number of VNFs in the $k$-th SFC. Hence, there are three SFCs in Fig. \ref{fig1}, i.e., $\mathcal{V}_1 =\{ v^1_1 \rightarrow v^2_1 \rightarrow v^3_1 \}$, $\mathcal{V}_2 =\{ v^1_2 \rightarrow v^2_2 \rightarrow v^3_2 \}$, and $\mathcal{V}_3 =\{ v^1_3 \rightarrow v^2_3 \rightarrow v^3_3 \}$. It is worth mentioning that, as shown in Fig. \ref{fig1}, there exists a control center on the ground. It contains the NFV management and orchestration (NFV MANO) to help implement and manage on the UAV and satellite nodes. Moreover, there are distributed NFV MANOs on the UAV and satellite nodes, so that each node can quickly judge and transmit tasks and information.

The VNFs of SFCs can be deployed on any UAV or satellite node. However, in reality, nodes may fail due to equipment failure, hardware damage, etc. Hence, in this paper, we mostly consider the node failure, resulting in the failure of VNF processing, and the corresponding failure of links connected to failed nodes. Therefore, it is necessary to redeploy and schedule SFCs on these failed nodes and links for service recovery. As shown in Fig. \ref{fig2}(a), there exist one UAV and one satellite node malfunctioning, preventing transmission from links connected to them. Hence, VNFs originally deployed on these two nodes need to be redeployed. As shown in Fig. \ref{fig2}(b), the two affected SFCs search for available UAVs and satellites and redeploy VNFs near the failed nodes. Consequently, the original deployment paths are abandoned and new paths are obtained. SFCs are able to successfully complete the deployment tasks. Therefore, we focus on quickly finding suitable available nodes and links for redeployment in the face of resource failure. The entire process of resource failure and SFC redeployment is shown in Fig. \ref{fig:flow diagram}.

\subsection{Channel Model}
\subsubsection{Channel Model of G2U, U2G, U2U, and S2G}
There exist six types of channels. Among them, the maximum data rate of G2U, U2G, U2U, and S2G is expressed according to Shannon formula \cite{6773024}, i.e.,
\begin{equation}
    R_{(i^t,j^t)}^n=B\mathrm{log}{}_{2}(1+\Omega^t ), \forall (i^t,j^t) \in \mathcal{Y}_l \setminus \{\mathcal{Y}_{us} \cup \mathcal{Y}_{ss} \},
\end{equation}
where $R_{(i^t,j^t)}^n=R_{(i^t,j^t)}^{gu} \cup R_{(i^t,j^t)}^{ug} \cup R_{(i^t,j^t)}^{uu} \cup R_{(i^t,j^t)}^{sg}$, and $B=B_{gu} \cup B_{ug} \cup B_{uu} \cup B_{sg}$ represent the bandwidth of G2U, U2G, U2U and S2G, respectively. $\Omega^t =\Omega_{gu}^t \cup \Omega_{ug}^t \cup \Omega_{uu}^t\cup \Omega_{sg}^t$ indicate the signal-to-noise ratio (SNR) of G2U, U2G, U2U and S2G, respectively. 

In detail, the communication link between the UAV and the ground is line-of-sight \cite{9714482}. Hence, the SNR of G2U and U2G can be expressed as
\begin{equation}
    \Omega_{gu}^t = \frac{P^{tr}G_0}{\sigma _0^2(d^t_{gu})^2}=\frac{P^{tr}G_0}{\sigma _0^2[(a^t_u-a_g)^2+(b^t_u-b_g)^2+h^2_u]^2},
\end{equation}
where $P^{tr}= P^{tr}_g \cup P^{tr}_u$ indicates the transmission power of the ground and the UAV, respectively. $G_0$ denotes the reference channel gain when the distance between the UAV and the ground station $d^t_{gu}=1$. $\sigma _0^2$ is the White Gaussian noise power. $\{(a^t_u,b^t_u),(a^t_g,b^t_g)\}$ represents the coordinates of the UAV and the ground station, respectively. $h^t_u$ indicates the flight height of the UAV. 

Following \cite{8424236}, the U2U channel can be regarded as a free-space propagation scenario, and the SNR of U2U is
\begin{equation}
    \Omega_{uu}^t = \frac{P_{uu}}{\sigma _{uu}^2}10^{\frac{-L^t_{uu}}{10}},
\end{equation}
where $P_{uu}$ is the transmission power, and $\sigma _{uu}^2$ denotes the noise power between the two UAVs. $L_{uu}^t$ indicates the path loss of U2U, i.e.,
\begin{equation}
    L^t_{uu} = 20\log_{10} d^t_{uu} + 10\log_{10}\left(\frac{4\pi f_{uu}}{c}\right)^2,
\end{equation}
where $d^t_{uu}$ represents the distance between two UAVs. $f_{uu}$ indicates the carrier frequency, and $c$ is the light speed.

The S2G channels are influenced by atmospheric precipitation. Hence, the state of S2G channels can be predicted using the meteorological satellites \cite{8353906}, so that the SNR of S2G is calculated as
\begin{equation}
    \Omega_{sg}^t = \frac{P_{sg}G_{sg}^{tr}G_{sg}^{re}L_e\alpha^tL_s^t }{N_0B_{sg}},
\end{equation}
where $B_{sg}$ denotes the bandwidth of S2G, and $N_0$ is the reference noise. $P_{sg}$ indicates the transmission power. $G_{sg}^{tr}$ and $G_{sg}^{re}$ represent the transmitter antenna gain of the satellite and the receiver antenna gain of the ground station, respectively \cite{10360859}, \cite{balanis1996antenna}. $L_e$ denotes the slant-path length , and $\alpha^t$ is the attenuation per kilometer in time slot $t$ \cite{ITU-R-P618-14}. $L_s^t$ indicates the related free space loss, i.e.,
\begin{equation}
    L_s^t = \left(\frac{c}{4\pi S^t f^{cen}}\right)^2,
\end{equation}
where $S^t$ represents the maximum slant range in time slot $t$, and $f^{cen}$ is the centering frequency.

\subsubsection{Channel Model of U2S and S2S}
For the remaining two channels of U2S and S2S, the maximum data rate \cite{TheFederatedSatellite} is obtained as
\begin{equation}
    R^s_{(i^t,j^t)}=\frac{P_{ij}G_{ij}^{tr}G_{ij}^{re}L_s^tL_l}{(E_b/N_0)_{r}k_BT_sS_m}, \forall (i^t,j^t) \in \mathcal{Y}_{us} \cup \mathcal{Y}_{ss},
\end{equation}
where $R^s_{(i^t,j^t)}=R^{us}_{(i^t,j^t)} \cup R^{ss}_{(i^t,j^t)}$. $P_{ij}$, $G_{ij}^{tr}$ and $G_{ij}^{re}$ indicate the transmission power, the transmitting gains, and the receiving antenna gains, respectively. $L_l$ represents the total line loss, and $(E_b/N_0)_{r}$ is the ratio of the received energy per bit to noise density. $k_B$, $T_s$ and $S_m$ denote the Boltzmann constant, the system noise temperature, and the maximum slant range, respectively.

\subsection{Energy Model}
\subsubsection{Energy Cost of UAV}
The total energy consumption of UAVs includes the path energy cost $E_{i^t,u}^p$ and the communication energy cost $E_{i^t,u}^c$, i.e.,
\begin{equation}
E_{i^t,u}=E_{i^t,u}^p+E_{i^t,u}^c,\forall i^t\in\mathcal{E}_{u},t\in T, \label{formula:energycost of HAP}
\end{equation}
in which
\begin{equation}
    E_{i^t,u}^p=P^m_{i^t} \frac{\Vert p _{i^t}-p _{i^{t+1}}\Vert _2}{s_{i^t}}+P ^h_{i^t}\cdot \tau, \forall i^t\in\mathcal{E}_{u},t\in T,
\end{equation}
where $p_{n^t_i}$ is the position coordinate of UAV $i^t$. $P^m_{i^t}$ and $P ^h_{i^t}$ denote the moving and hovering power \cite{9865119}, respectively, i.e.,
\begin{equation}
    P^m_{i^t}=\frac{s_{i^t}}{s^{\rm max}_{i^t}}(P^{\rm max}_{i^t}-P^{h}_{i^t}), \forall i^t\in\mathcal{E}_{u},t\in T,
\end{equation}
where $s_{i^t}$ and $s^{\rm max}_{i^t}$ is the moving speed, and the maximum speed of UAV $i^t$. $P^{\rm max}_{i^t}$ denotes the power when UAV $i^t$ is at the maximum speed, and $P^{h}_{i^t}$ represents the hovering power,
\begin{equation}
    P^{h}_{i^t} = \sqrt{\frac{(g M_{i^t})^3}{2 \pi \vartheta \iota^2_{i^t} \kappa_{i^t} }} ,\forall i^t\in\mathcal{E}_{u},t\in T,
\end{equation}
where $g$ denotes the earth gravity acceleration, and $M_{i^t}$ is the mass of UAV $i^t$. $\vartheta$, $\iota_{i^t}$, and $\kappa_{i^t}$ indicate the air density, radius, and the number of propellers in UAV $i^t$, respectively. Moreover, the communication energy cost can be expressed as
\begin{equation}
    E_{i^t,u}^c=\underset{k\in\mathcal{K}}{\sum}\underset{j^t \in\mathcal{E}}{\sum}\frac{P^{tr}_{i^t}y^k_{(i^t,j^t)} \Delta_k}{R^u_{(i^t,j^t) }}, \forall i^t\in\mathcal{E}_{u},t\in T,
\end{equation}
where $P^{tr}_{i^t}$ and $\Delta_k$ represent the transmitting power and the data amount of SFC $\mathcal{V}_k$, respectively. $y^k_{(i^t,j^t)}=1$ indicates SFC $\mathcal{V}_k$ passes through link $(i^t,j^t)$, and otherwise $y^k_{(i^t,j^t)}=0$. $R^u_{(i^t,j^t) }$ is composed of $R^{gu}_{(i^t,j^t) }$ and $R^{uu}_{(i^t,j^t) }$, denoting the maximum data rate of G2U and U2U, respectively.

\subsubsection{\textcolor{black}{Energy Cost of Satellite}}
The total energy consumption of satellites includes the reception energy cost $E_{i^t,s}^r$, transmission energy cost $E_{i^t,s}^t$, and general operation energy cost $E_{i^t,s}^o$, respectively, i.e.,
\begin{equation}
E_{i^t,s}=E_{i^t,s}^r+E_{i^t,s}^t+E_{i^t,s}^o, \forall i^t\in\mathcal{E}_{s},t\in T.
\end{equation}
In detail, we have: $\forall i^t\in\mathcal{E}_{s},t\in T$,
\begin{equation}
    E_{i^t,s}^{r}\!=\!\underset{k\in\mathcal{K}}{\sum}\Delta_k\!\left(\underset{j^t \in\mathcal{E}_{u}}{\sum}\!\!\frac{P_{us}^{re}y^k_{(j^t,i^t)} }{R^{us}_{(j^t,i^t) }}\!+\!\!\!\underset{j^t\in\mathcal{E}_{s}}{\sum}\!\!\frac{P_{ss}^{re}y^k_{(j^t,i^t)}}{R^{ss}_{(j^t,i^t) }}\!\right)\!,\label{formula: Energycost LEO}
\end{equation}
and \vspace{-2mm}
\begin{equation}
    E_{i^t,s}^{t}\!=\!\underset{k\in\mathcal{K}}{\sum}\Delta_k\!\left(\underset{j^t \in\mathcal{E}_{s}}{\sum}\!\!\frac{P_{ss}^{tr}y^k_{(i^t,j^t)}}{R^{ss}_{(i^t,j^t) }}\!+\!\!\!\underset{j^t\in\mathcal{E}_{g}}{\sum}\!\!\frac{P_{sg}^{tr}y^k_{(i^t,j^t)}}{R^{sg}_{(i^t,j^t) }}\!\right)\!,
\end{equation}
where $P_{us}^{re}$ and $P_{ss}^{re}$ indicate the received power of U2S and S2S, respectively. $P_{ss}^{tr}$ and $P_{sg}^{tr}$ represent the transmitted power of S2S and S2G, respectively \cite{jia2021vnf}.

\section{Problem Formulation\label{sec:Problem-Formulation}}
\subsection{Constraints}
\subsubsection{SFC Deployment Constraints}
Firstly, for each VNF in the SFC, only one node can be selected to deploy in a time slot, i.e.,
\begin{equation}
    \underset{i^t \in \mathcal{E}\backslash \mathcal{E}_g}{\sum} \!\! x_{v^m_k,i^t}=1, \forall  v^m_k \in \mathcal{V} _k, \label{cons:x01}
\end{equation}
where the variable $x_{v^m_k,i^t}=1$ represents VNF $v^m_k$ is deployed on node $i^t$, and otherwise $x_{v^m_k,i^t}=0$. Besides, all SFCs need to complete the processing and deployment of all contained VNFs, i.e.,
\begin{equation}
    \underset{k \in \mathcal{K}}{\sum}\underset{v^m_k \in V_k}{\sum}\underset{i^t \in \mathcal{E}\backslash \mathcal{E}_g}{\sum}  x_{v^m_k,i^t}= \sum_{k=1}^{K} l_k.
\end{equation}
In addition, for each SFC $\mathcal{V}_k$, VNFs needs to be deployed sequentially, i.e., the subsequent VNF $v^{m+1}_k$ must be processed after completing the deployment of the previous VNF $v^m_k$,
\begin{equation}
    t_{v^{m+1}_k}-t_{v^{m}_k} \geq \underset{i^t\in \mathcal{E}\backslash \mathcal{E}_g}{\sum} x_{v^m_k,i^t} \cdot \sigma _{v_k^m}/{\varphi _{i^t}} , \forall  v^m_k\in \mathcal{V} _k,
\end{equation}
where $\sigma _{v_k^m}$ represents the computing resource consumed by VNF $v^m_k$, and $\varphi _{i^t}$ denotes the computation ability of node $i^t$. It is worth noting that a VNF can be deployed across time slots if it cannot be deployed in one time slot. Moreover, the flow constraints need to be satisfied, i.e.,
\begin{subnumcases} {\label{flowcons}}
    \underset{(o^t,j^t)\in \mathcal{Y}_l}{\sum}\!\!\!\! y^k_{(o^t,j^t)}=1,\forall k \in \mathcal{K}, o^t \in \mathcal{E}, t \in T, \label{cons:flowconsOD1}\\ 
    \underset{(i^t,j^t)\in \mathcal{Y}_l}{\sum} \!\!\!\!y^k_{(i^t,j^t)}+\!\!\!\!\!\!\underset{(j^{t-1},j^t)\in \mathcal{Y}_{t-1}}{\sum} \!\!\!\!\!\!\!y^k_{(j^{t-1},j^t)}\!=\!\!\!\! \underset{(j^t,i^t)\in \mathcal{Y}_l}{\sum} \!\!\!\!\! y^k_{(j^t,i^t)} +\nonumber\\
    \!\!\!\underset{(j^t,j^{t+1})\in \mathcal{Y}_t}{\sum} \!\!\!\!\!y^k_{(j^t,j^{t+1})}, \forall k\in \mathcal{K}, j^t \in \mathcal{E}, t \in \{2,...,T\!-\!1\}, \label{cons:flowconserS}\\
    \underset{(i^t,d^t)\in \mathcal{Y}_l}{\sum} \!\!\!\!y^k_{(i^t,d^t)}=1,\forall k \in \mathcal{K}, d^t \in \mathcal{E}, t \in T, \label{cons:flowconsOD2}
\end{subnumcases}
where $o^t$ and $d^t$ denote the origin node and destination node in SFC deployments, respectively. Binary variable $y^k_{(i^t,j^t)}=1$ indicates SFC $\mathcal{V}_k$ passes through link $(i^t,j^t)$, and otherwise $y^k_{(i^t,j^t)}=0$.

\subsubsection{Resource Constraints}
When each SFC is transmitted on the channel, the transmission resources consumed cannot exceed the capacity of the channel, i.e.,
\begin{equation}
    \underset{k \in \mathcal{K}}{\sum} y^k_{(i^t,j^t)} \Delta _k \leq R_{(i^t,j^t)}\cdot \tau, \forall (i^t,j^t) \in \mathcal{Y}_l, t\in T,
\end{equation}
where $R_{(i^t,j^t)}$ represents the maximum data rate of link $(i^t,j^t)$. In addition, when the SFC is processed on node $i^t$, the computing resources cannot exceed the computing capacity of the node $C_ {i^t}$, i.e.,
\begin{equation}
    \underset{k\in \mathcal{K} }{\sum} \underset{v^m_k\in \mathcal{V}_k}{\sum} x_{v^m_k,i^t} \sigma _{v_k^m}\leq C_ {i^t} ,\forall i^t \in \mathcal{E}\backslash \mathcal{E}_g, t \in T.
\end{equation}
When the SFC is not processed on the node and transmitted on the channel, the SFC selects to be stored on the node. However, the data amount stored on node $i^t$ cannot exceed the storage capacity of the node $A_{i^t}$, i.e.,
\begin{equation}
    \underset{k\in \mathcal{K}}{\sum}  z^k_{(i^t,i^{t+1})} \Delta_k  \leq A_{i^t},\forall (i^t,i^{t+1})\in \mathcal{Y}_t, t \in T,
\end{equation}
where $z^k_{(i^t,i^{t+1})}=1$ represents the SFC $\mathcal{V}_k$ is stored on node $i^t$ from $t$ to $t+1$, and otherwise, $z^k_{(i^t,i^{t+1})}=0$.

Furthermore, the total computation energy consumption on node $i^t$ and the energy consumed by the node operation $E^{T}_{i^t}$ cannot exceed the energy capacity $E^M_{i^t}$, i.e.,
\begin{equation}
    \underset{k\in \mathcal{K} }{\sum} \underset{v^m_k\in \mathcal{V}_k}{\sum} x_{v^m_k,i^t} \sigma _{v_k^m} e^c_{i^t}+E^{T}_{i^t}\leq E^{M}_{i^t} ,\forall {i^t} \in \mathcal{E}\backslash \mathcal{E}_g, t \in T, \label{cons:energy}
\end{equation}
where $e^c_{i^t}$ denotes the energy consumption per unit of computing resource on node $i^t$.

\subsection{Formulation}
The optimization problem is formulated to minimize the total time delay to process and redeploy all SFCs. For SFC $\mathcal{V}_k$, the time consumption $\mathcal{T}_k$ includes transmission delay $\mathcal{T}^r_k$, processing delay $\mathcal{T}^p_k$ , storage delay $\mathcal{T}^s_k$, delay caused by SFC  redeployment, and delay caused by the redeployment of other SFCs $\mathcal{T}^o_k$, i.e.,
\begin{equation}
    \mathcal{T}_k = \mathcal{T}^r_k + \mathcal{T}^p_k + \mathcal{T}^s_k + \underset{v^m_k\in \mathcal{V}_k}{\sum} w^m_k \mathcal{T}^{re}_{v^m_k} + \mathcal{T}^o_k, \forall k \in \mathcal{K},
\end{equation}
where
\begin{equation}
    \mathcal{T}^r_k=\underset{i^t \in \mathcal{E}\backslash \mathcal{E}_g}{\sum}\underset{v^m_k\in \mathcal{V}_k}{\sum} x_{v^m_k,i^t} \sigma _{v_k^m} / \varphi _{i^t}, \forall k \in \mathcal{K},
\end{equation}
\begin{equation}
    \mathcal{T}^p_k=\underset{(i^t,j^t)\in \mathcal{Y}_l}{\sum} y^k_{(i^t,j^t)}\Delta_k / R_{(i^t,j^t)}, \forall k \in \mathcal{K},
\end{equation}
and
\begin{equation}
    \mathcal{T}^s_k=\underset{(i^t,i^{t+1})\in \mathcal{Y}_t}{\sum}z^k_{(i^t,i^{t+1})}, \forall k \in \mathcal{K},
\end{equation}
in which $w^m_k$ indicates the redeployment of VNF $v^m_k$. In detail, $w^m_k=1$ represents VNF $v^m_k$ of SFC $\mathcal{V}_k$ needs to be redeployed, and otherwise $w^m_k=0$. Let $\mathcal{T}^{re}_{v^m_k}$ denote the delay generated after the redeployment of VNF $v^m_k$. Then, the optimization problem can be expressed as
\begin{equation}{\label{optimal}}
    \begin{aligned}
    \mathscr{P}0:\;&\underset{\boldsymbol{X},\boldsymbol{Y},\boldsymbol{Z},\boldsymbol{W}}{\textrm{min}}\; \sum_{k \in \mathcal{K}}\mathcal{T}_k \\
    &\begin{array}{r@{}l@{\quad}l}
        \textrm{s.t.} \;\;&(\ref{cons:x01})-(\ref{cons:energy}), \\
             &x_{v^m_k,i^t}, y_{(i^t,j^t)}^k, z^k_{(i^t,i^{t+1})}, w^m_k\in \{0,1\},\\
        \end{array}
    \end{aligned}
\end{equation}
where $\boldsymbol{X}=\{x_{v^m_k,i^t},\forall k\in \mathcal{K}, i^t \in \mathcal{E}_u \cup \mathcal{E}_s, t\in T\} $, $\boldsymbol{Y}=\{y^k_{(i^t,j^t)},\forall k\in \mathcal{K}, (i^t,j^t) \in \mathcal{Y}_l, t\in T\}$, $\boldsymbol{Z}=\{z^k_{(i^t,i^{t+1})},\forall k\in \mathcal{K}, (i^t,i^{t+1})\in\mathcal{Y}_t, t\in T\}$, and $\boldsymbol{W} =\{w^m _k,\forall v^m_k \in \mathcal{V}_k, k\in \mathcal{K}\}$. We note that $\mathscr{P}0$ is an ILP problem, which is NP-hard and intractable to figure out in finite time complexity \cite{wolsey1999integer}. Therefore, we design the algorithm based on matching game.

\section{algorithm design\label{sec: DRL algorithm}}
In this section, we deal with $\mathscr{P}0$ in the case of resource failure, mainly based on the two-sided matching game \cite{doi:10.1287/mnsc.47.9.1252.9784}. The entire algorithm of SFC redeployment for resource failure is described in Algorithm \ref{FRMG-SAGIN}, termed as Failure and Recovery by Matching Game in SAGIN (FRMG-SAGIN). It includes the initialization (line \ref{1line1}), SFC deployment under normal circumstances (lines \ref{1line3}-\ref{1line8}), and SFC recovery in the case of resource failure (line \ref{1line11}). When SFCs are not affected by failed nodes, they are deployed according to initialized paths (line \ref{1line5}). When multiple SFCs arrive simultaneously at the same node, and are prepared to be processed, these SFCs are sorted in an ascending order based on the data amount. Then, the node selects SFCs sequentially until its capacity limitation (line \ref{1line7}). The remaining SFCs wait and are stored at the node, to be assessed and processed in the next time slot (line \ref{1line8}). If SFCs are affected by node failure, the resource service recovery and SFC redeployment are completed through Algorithm \ref{Match}. When all SFCs are deployed, the total time consumption by all SFCs is calculated.

\begin{algorithm}[!t]
    \caption{FRMG-SAGIN\label{FRMG-SAGIN}}

    \begin{algorithmic}[1]
        \REQUIRE The positions of UAV and satellite nodes, starting points of SFCs, destination points of SFCs, the data amount $\Delta_k$ of SFCs.
        \ENSURE The time consumption for the processing and redeployment of all SFCs.
        \STATE Initialize the SFC deployment by Algorithm \ref{Initialization}. \label{1line1}\\
        \FOR {each time slot $t$} \label{1line2}
            \IF {there exist SFCs do not complete deployments} \label{1line3}
                \IF {SFCs are not affected by node failure} \label{1line4}
                    \STATE Deploy SFCs according to the path obtained from initialization. \label{1line5}
                    \IF {multiple SFCs arrive at the same node $i^t$} \label{1line6}
                        \STATE Select SFCs according to $\Delta_k$  in an ascending order until the node capacity is full.\label{1line7} \\
                        \STATE The remaining SFCs wait and store on $i^t$ to $i^{t+1}$.\label{1line8} 
                    \ENDIF
                \ELSE 
                    \STATE Redeploy the affected SFCs by Algorithm \ref{Match}.\label{1line11} \\
                \ENDIF
            \ELSE
                \STATE Sum time consumption of all SFCs.\label{1line14} 
            \ENDIF
        \ENDFOR
    \end{algorithmic}
\end{algorithm}

\subsection{Algorithm for SFC Initialization}
As described in Algorithm \ref{Initialization}, we initialize the SFC deployment, without considerations of node failures. The first step is to obtain the initial position of UAV and satellite nodes, as well as the data amount, starting and ending nodes of SFCs. For each SFC, the closest UAV nodes to its starting and ending points are found. Firstly, we set the distance from the starting node $i_s$ to itself as 0, and to all other nodes as infinity. A collection $\mathbb{C}_{uv}$ for recording unvisited nodes is created (line \ref{2line3}). Then, we calculate the distance from $i_s$ to nodes in $\mathbb{C}_{uv}$, and set the node with the shortest distance $d_{sc}$ as the current node $i_c$ (line \ref{2line5}). For each neighbor node $i_n$ of $i_c$, the distance $d_{new}$ from node $i_s$ to $i_n$ passing the current node $i_c$ is calculated (line \ref{2line10}). If $d_{new}$ is shorter than $d_{sc}$, the path and distance of the neighbor node are updated. The preceding steps are repeated until all nodes are accessed (lines \ref{2line6}-\ref{2line13}). At last, the set is obtained including the shortest path from the start node to the destination node.

\begin{algorithm}[!t]
    \caption{Initialization of SFC Deployments\label{Initialization}}

    \begin{algorithmic}[1]
        \REQUIRE The positions of UAV and satellite nodes, starting points of SFCs, destination points of SFCs, the data of SFCs, and the set of neighbor nodes $\mathbb{C}_{uv}$.
        \ENSURE The shortest paths for SFC deployments.
        \STATE \textbf{Initialization:} the null set of shortest distances from the starting point $i_s$ to other nodes, and the null set of neighbor nodes $\mathbb{C}_{uv}$. \label{2line1} \\
        \STATE Set the distance from node $i_s$ to $i_s$ as 0, and from $i_s$ to other nodes as infinity. \label{2line2}\\
        \STATE Put all nodes except $i_s$ into $\mathbb{C}_{uv}$.\label{2line3}\\
        \FOR {\rm{$\mathbb{C}_{uv}$ is not empty}}\label{2line4}
            \STATE $i_c \leftarrow$ node in $\mathbb{C}_{uv}$ with minimum distance $d_{sc}$ from $i_s$.\label{2line5}\\
            \FOR {all neighbor nodes $i_n$ of $i_c$}\label{2line6}
                \IF {$i_n$ \rm{is not in} $\mathbb{C}_{uv}$}\label{2line7}
                    \STATE Break.\label{2line8}
                \ENDIF\label{2line9}
                \STATE $d_{new} \leftarrow$ the distance from $i_s$ to $i_n$ passing $i_c$.\label{2line10}
                \IF {$d_{new} < d_{sc} $}\label{2line11}
                    \STATE $d_{sc} \leftarrow d_{new}$.\label{2line12}
                \ENDIF\label{2line13}
                \STATE Remove $i_c$ from $\mathbb{C}_{uv}$.\label{2line14}
            \ENDFOR
        \ENDFOR
    \end{algorithmic}
\end{algorithm}

\subsection{Matching Game based Robust Service Recovery}
When a node fails, all links connected to the node also fail. Therefore, the SFCs transmitting on these links can only return to the previous node to find the next available node. Hence, we design a matching game based algorithm for the recovery of these SFCs to be redeployed, as described in Algorithm \ref{Match}.


In the matching game, there are two parties of players. One party of players sends out a match request, and the other party chooses to accept or reject the match request. The final desired result is that all players are satisfied with their matched opponents. The match reaches a stable state, and no two players are willing to break the current matched results and form a new match pair \cite{8869709}. Each party has a preference list of all players in the other party, i.e., who it prefers to match. In this paper, SFCs are set as the matching proposers, and nodes are set as the matching receivers. Hence, there are two preference lists for there two matching parties, i.e., SFCs and nodes, respectively. 

Firstly, we build the preference lists for SFCs and nodes (line 3). Since the optimization problem, i.e., the completing time of SFC deployment should be as short as possible, the preference list $L_S$ of an SFC is the order of time consumed from the current node to the next node $i^t_A$, i.e., 
\begin{equation}
    L_S=T^r_{cA} + T^r_{Ad}, \label{pre1}
\end{equation}
where $T^r_{cA}$ indicates the time consumed by the distance between the current node $i^t_c$ and $i^t_A$. $T^r_{Ad}$ represents the time consumed by the shortest distance between $i^t_c$ and the destination node $i^t_d$. It is worth noting that since (\ref{pre1}) is related to the time consumed by the path, the precondition for a node to be contained in the preference list is that the current node and the next node can communicate and transmit SFCs.

\begin{algorithm}[!t]
    \caption{Matching Game based Robust Service Recovery\label{Match}}

    \begin{algorithmic}[1]
        \REQUIRE All SFCs required redeployment $\mathcal{V}_k$, and all available nodes $i^t_A \in \mathcal{E}_u \cup \mathcal{E}_s$.
        \ENSURE Stable matching between SFCs and nodes.
        \STATE \textbf{Initialization:} the unmatched list $L_U$ with all SFCs required redeployment.\label{3line1}
        \FOR {each time slot $t$}\label{3line2}
            \STATE Obtain preference lists $L_S$ and $L_N$.\label{3line3}\\
            \FOR {$L_U$ is not empty}\label{3line4}
                \STATE Choose an SFC $\mathcal{V}_c$ in $L_U$ randomly.\label{3line5}\\
                \STATE Select the node $i^t_n$ at the front of $L_S$.\label{3line6}\\
                \IF {the capacity of $i^t_n$ is insufficient}\label{3line7}
                    \STATE Redeploy $\mathcal{V}_c$ on $i^t_n$.\label{3line8}\\
                    \STATE Remove $\mathcal{V}_c$ from $L_U$.\label{3line9}\\
                \ELSE\label{3line10}
                    \STATE Select the allocated SFC $\mathcal{V}_l$ with the lowest ranking in $L_N$.\label{3line11}\\
                    \IF {$\mathcal{V}_c \succ_{i^t_n} \mathcal{V}_l$}\label{3line12}
                        \STATE Replace $\mathcal{V}_l$ with $\mathcal{V}_c$ in the matching list of $i^t_n$.\label{3line13}\\
                        \STATE Update $L_U$.\label{3line14}
                    \ENDIF
                \ENDIF
            \ENDFOR 
        \ENDFOR
    \end{algorithmic}
\end{algorithm}

In order to minimize the total delay in the optimization problem and reduce the waste of time, nodes tend to choose SFCs that are being processed on the node and SFCs that are already waiting on the node. Hence, the preference list $L_N$ of a node is based on the sort of SFCs that select the node, which is expressed as 
\begin{equation}
    L_N=a \cdot \varsigma^p_k + b\cdot \varsigma ^s_k + c / \Delta _k, \label{nodeprelist}
\end{equation}
where $a$, $b$, and $c$ denote the weighted parameters, in which the value of $a$ is the largest and $c$ the smallest. $\varsigma^p _k=1$ indicates SFC $\mathcal{V}_k$ is being processed, and otherwise $\varsigma ^p_k=0$. $\varsigma ^s_k=1$ represents SFC $\mathcal{V}_k$ is being stored, and otherwise $\varsigma ^s_k=0$. The specific explanation of $L_N$ is that SFCs that are being processed ranks firstly, and SFCs stored on the node are lined up behind them. Other SFCs are sorted in an ascending order according to data amounts $\Delta _k$. In each loop, an unmatched SFC $\mathcal{V}_c$ is randomly chosen (line \ref{3line5}). According to the preference list $L_S$, node $i^t_n $ at the front of $L_S$ is selected (line \ref{3line6}). If the capacity of $i^t_n $ is insufficient, $\mathcal{V}_c$ is allocated to $i^t_n $ (lines \ref{3line7}-\ref{3line8}). Otherwise, the SFC $\mathcal{V}_l$ with the lowest rank in the preference list of $i^t_n$ from the current matching list of $i^t_n$ is selected, and compared with $\mathcal{V}_c $ (lines \ref{3line11}-\ref{3line12}). If $\mathcal{V}_c \succ_{i^t_n} \mathcal{V}_l$, where the symbol $\succ_{i^t_n}$ indicates that $\mathcal{V}_c$ ranks higher than $\mathcal{V}_l $ in $L_N$ of $i^t_n$, the matching list of $i^t_n$ adds $\mathcal{V}_c$ and deletes $\mathcal{V}_l $ (lines \ref{3line13}-\ref{3line14}). The loops end until all SFCs that need redeployment are matched with nodes or the capacity of nodes are exhausted. Thus, the matching between SFCs that need to be redeployed and nodes are obtained. SFCs can be successfully transmitted to the matched node through the corresponding link in the current time slot.

\subsection{Complexity Analysis of FRMG-SAGIN}
Assuming that there are $N$ nodes, $E$ links, and $M$ SFCs, the computational complexity is divided into two parts. The node with the smallest distance is found each time. The time complexity of this process is $\mathcal{O}(logN)$. When the distance of a node is updated, all the neighbors of the node and links need to be gone through \cite{gallo1988shortest}. Hence, the total time complexity is $\mathcal{O}((N+E)logN)$. In the matching process, each SFC requests the corresponding node for transmission according to the preference list, and the node compares and selects all SFCs that send transmission requests to it \cite{benedek2023complexity}. If the number of SFCs is $M$, the time complexity of the whole matching process is $\mathcal{O}(MN)$. Then, the total complexity is $\mathcal{O}((N+E)logN+MN)$. It is noted that the proposed algorithm has good applicability in small-scale networks, which can balance the network delay and the computational overhead.

\begin{table}[!t]
    \renewcommand\arraystretch{1.3}
	\begin{center}
		\caption{PARAMETER SETTING} \label{parameter setting}
		\begin{tabular}{|c|c||c|c|}
			\hline
			Description & Value & Description & Value \\
			\hline
            \hline
			$P^{tr}_{g}$ & 0.5W &  $P^{tr}_u$ & 10W \\
            \hline 
            $h_u$ & 100m & $P_{uu}$ & 10W\\
            \hline
            $f_{uu}$ & 2.4GHz & $\sigma_{uu}^2$ & $4\times 10^{-13}$W\\
            \hline
            $P_{sg}$ & 20W & $G_{sg}^{tr}G_{sg}^{re}$ & 42dB\\
            \hline
            $f^{cen}_{us}$ & 3.4GHz & $P_{us}$ & 10W\\
            \hline
            $P_{ss}$ & 20W &  $G_{us}^{tr}G_{us}^{re}$ & 42dB\\
            \hline 
            $G_{ss}^{tr}G_{ss}^{re}$ & 52dB &$T_s$ & 1000K\\
            \hline
            $L_l$ & 2dB & $N_0$ & -114dBm\\
            \hline
            $B_{gu}$ & 2MHz & $B_{uu}$ &4MHz \\
            \hline
            $B_{us}$ & 50MHz & $B_{ss}\&B_{sg}$ & 80MHz\\
            \hline
            $f^{cen}_{ss}$ & 2.2GHz & $f^{cen}_{sg}$ & 20GHz\\
            \hline
            $s_{i^t}$ & 12m/s & $P^{\rm max}_{i^t}$ & 5W\\
            \hline
            $\iota _{i^t}$ & 20cm & $\kappa_{i^t}$ & 4\\
            \hline
		\end{tabular}
	\end{center}
\end{table}

\section{Simulation Results\label{sec:Simulation-Results}}
\subsection{\textcolor{black}{Simulation Setups}}

In this section, we conduct simulations using Python to validate the robust recovery and redeployment of SFCs in the event of resource failure. Each  satellite and UAV maintains a quasi-static state during a time slot, whose length is 5 seconds. A total of 30 UAVs randomly fly within a square area of 2km $\times$ 2km. The initial distributions of UAVs is known. To avoid collisions, the distance between any two UAVs must be at least 20m, and if the distance exceeds 500m, these two UAVs are programmed not to communicate with each other. Additionally, two satellites obtain the position information from the Starlink G8-8 at approximately 9:40 p.m. on June 27, 2024, located at 32$^\circ$N, 119$^\circ$E. The movement trajectory of UAVs are clearly known, and the trajectories of satellites are regular. There exist a range of SFCs in total, each comprising 2 or 3 VNFs, with data amounts ranging from 200 Mbit to 800 Mbit. We  randomly update failed UAV and satellite nodes every 3 time slots, and the number is based on the Poisson distribution, with parameter of $\lambda$. Besides, the specific parameters used in simulations are listed in Table \ref{parameter setting}. During each simulation, we carry out 500 loops, where the initial position of UAVs, as well as the randomly generated data amount of SFCs, the number of VNFs, and the corresponding starting and ending coordinates are fixed in each loop. Moreover, in a loop, 500 internal loops are carried out, where the information of failed nodes is randomly generated according to the interval of time slots.

\subsection{\textcolor{black}{Simulation Results and Analyses}}
\subsubsection{Influences of different parameters}

\begin{figure}[!t]
	\centering
	{\includegraphics[width=.95\columnwidth]{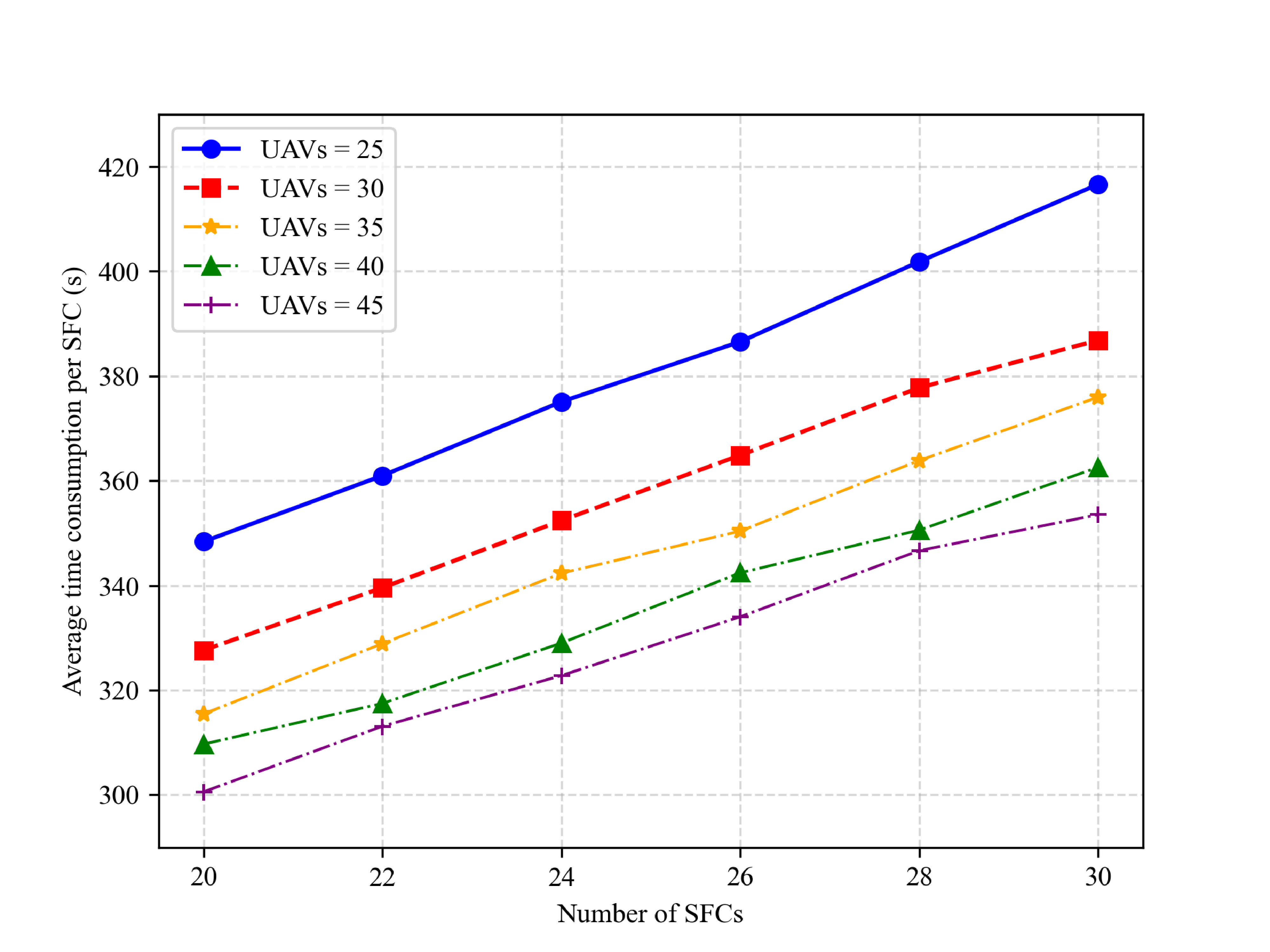}}
	\caption{The time consumption with different numbers of UAVs.}
    \label{fig:uavchange}
\end{figure}

\begin{figure}[!t]
	\centering
	{\includegraphics[width=.95\columnwidth]{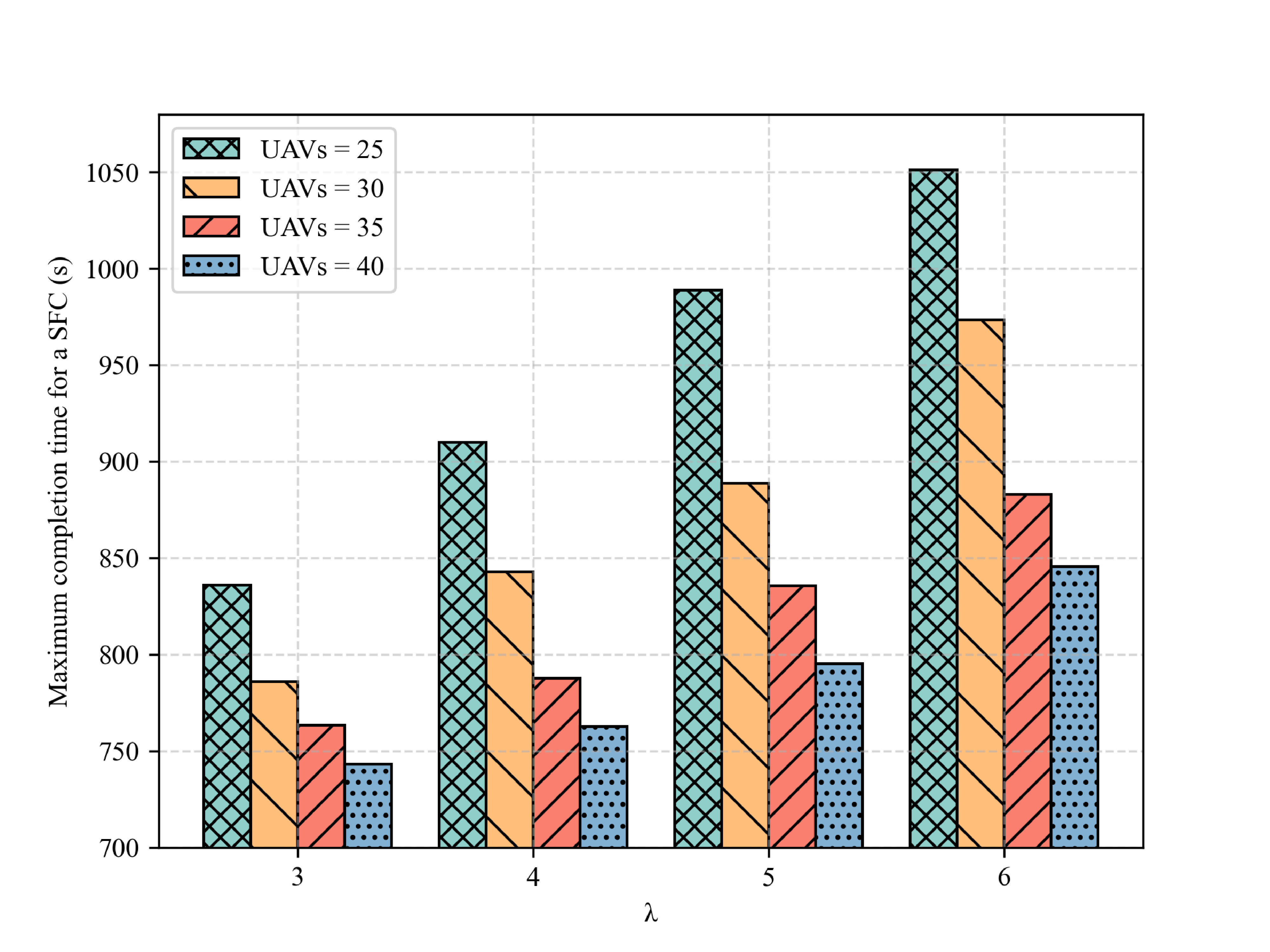}}
	\caption{Maximum completion time for an SFC under different numbers of failed nodes.}
    \label{fig:nodechange_timeslot}
\end{figure}

\begin{figure*}[!t]
	\centering
	\subfloat[]{\includegraphics[width=.95\columnwidth]{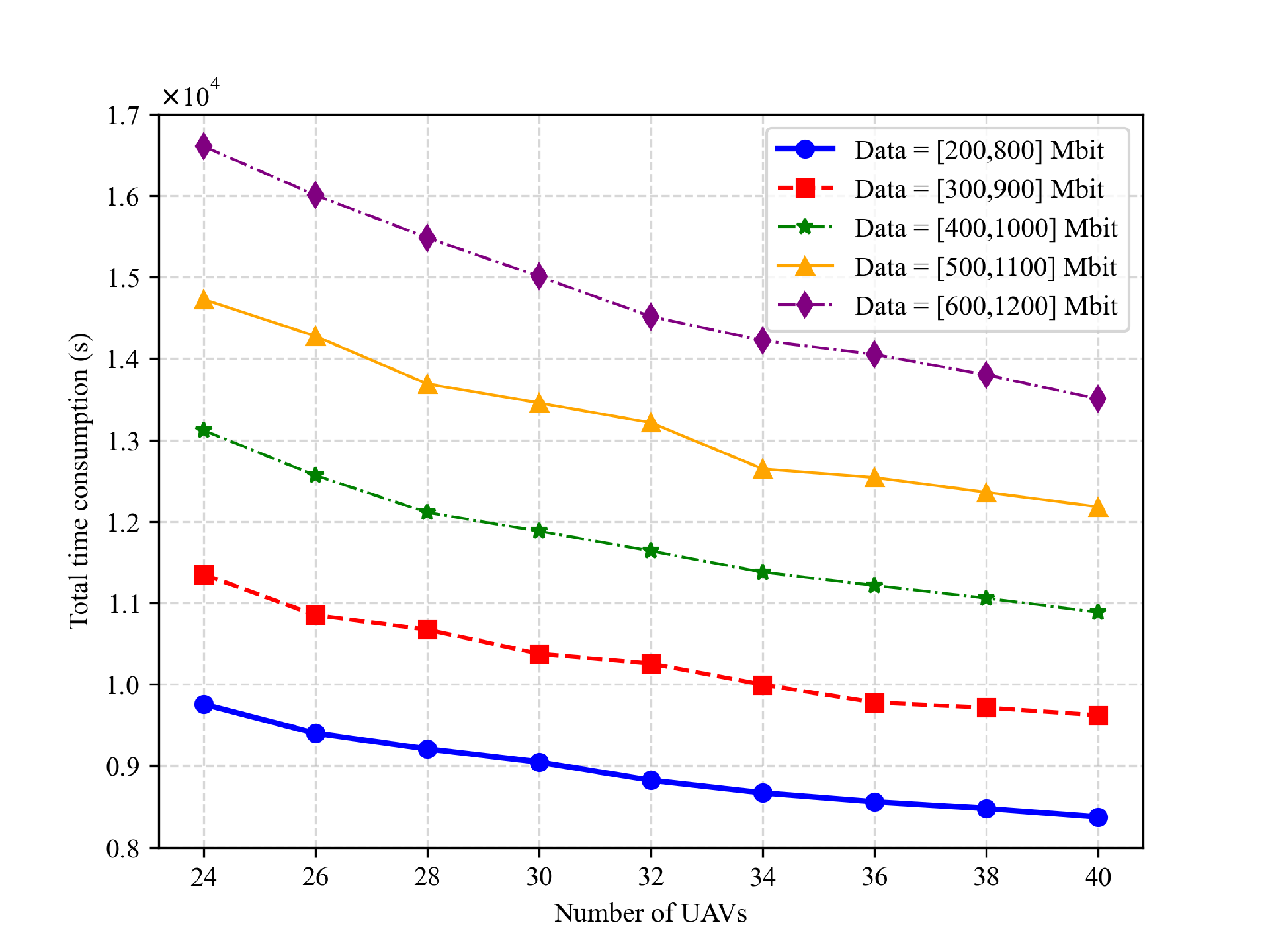}}\label{fig:datachange} \hspace{7mm}
    \subfloat[]{\includegraphics[width=.95\columnwidth]{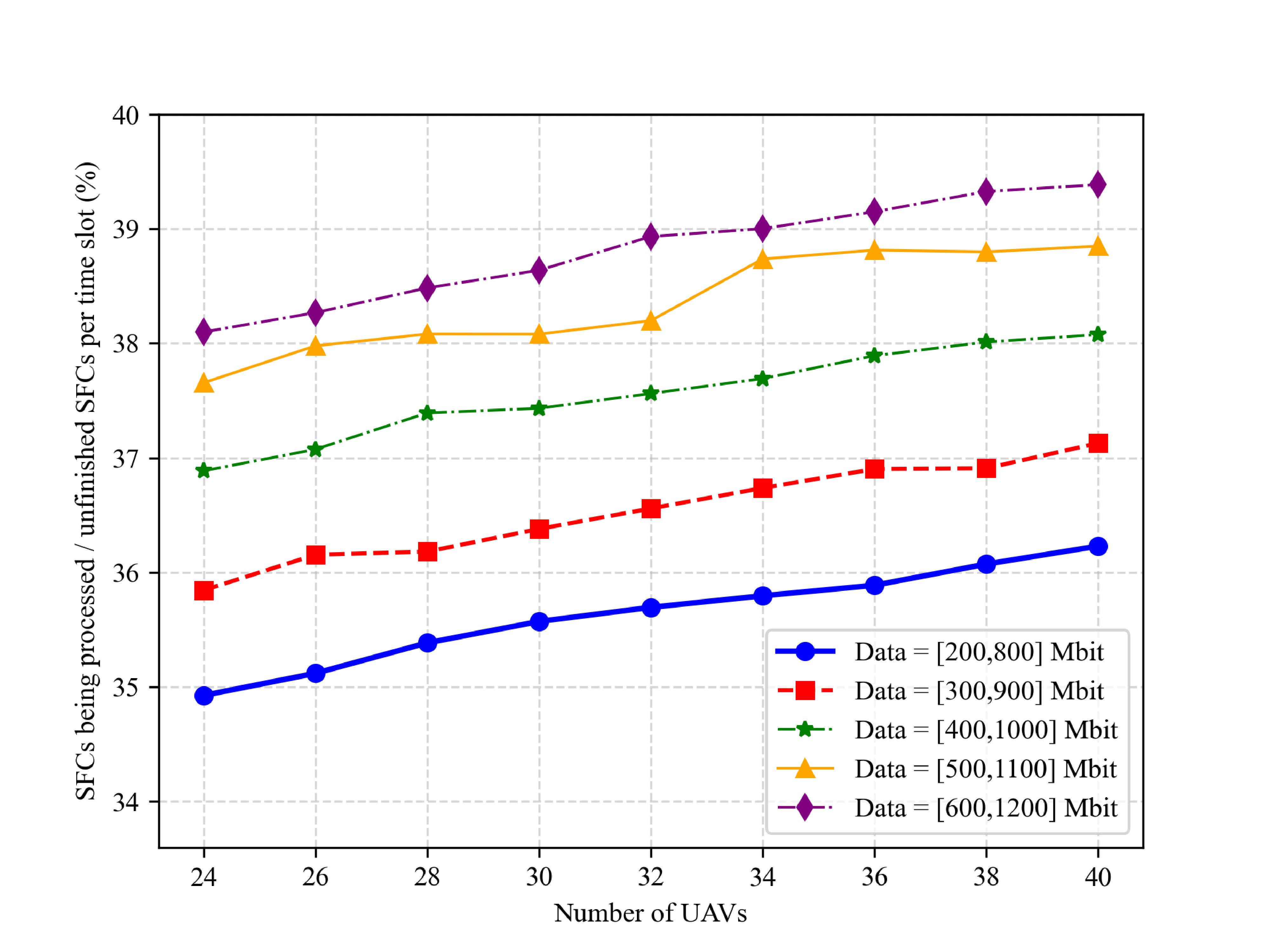}} \label{fig:dataprocess}
	\caption{Comparison with different intervals of data amounts of SFCs. (a) Total time consumption. (b) Ratio of SFCs being processed to unfinished SFCs per time slot.}
    \label{fig:dataamount}
\end{figure*}

\begin{figure}[!t]
	\centering
	{\includegraphics[width=.95\columnwidth]{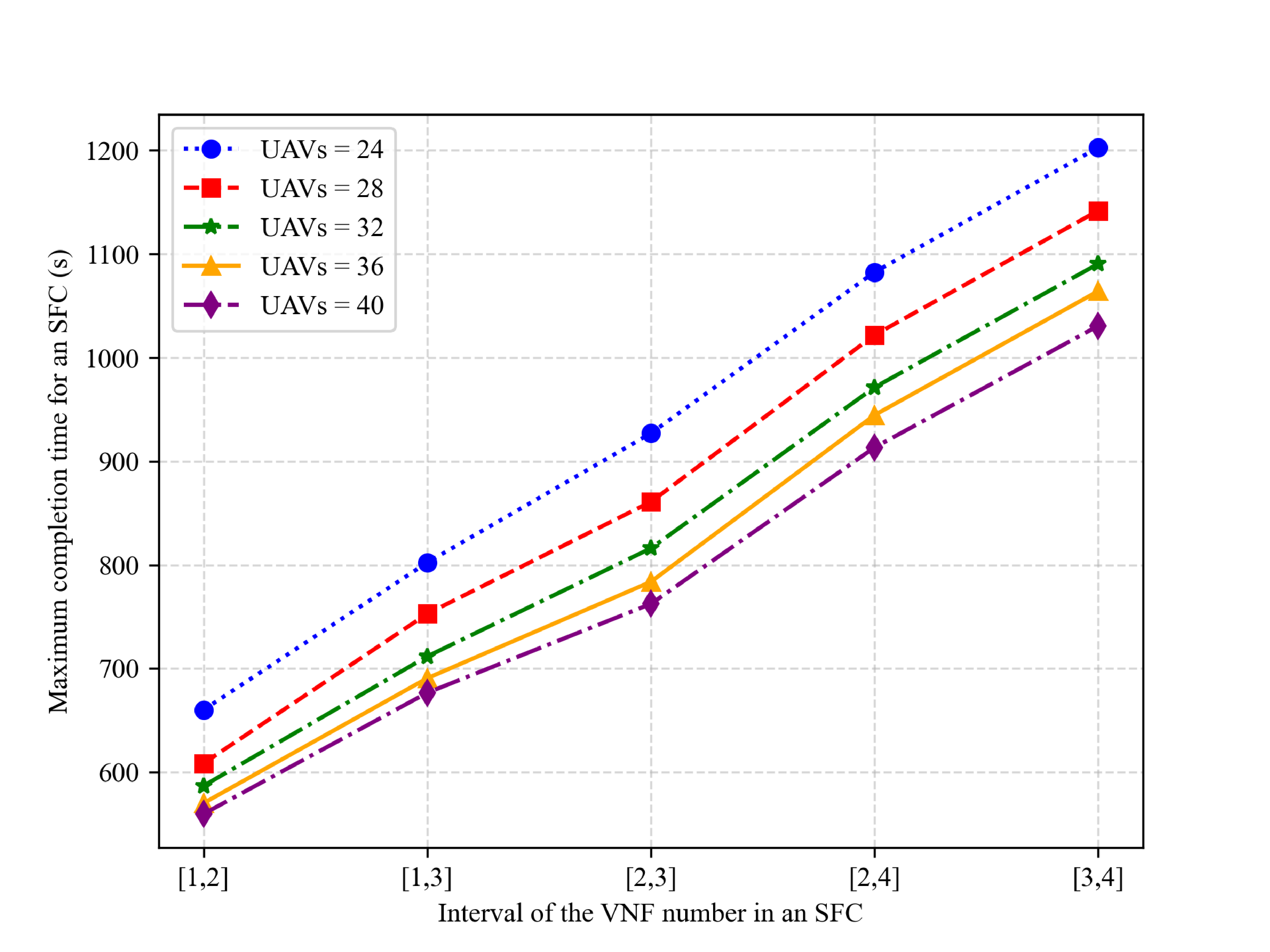}}
	\caption{Maximum completion time for an SFC with different numbers of VNFs in an SFC.}
    \label{fig:vnfchange}
\end{figure}

\begin{figure}[!t]
	\centering
	{\includegraphics[width=.95\columnwidth]{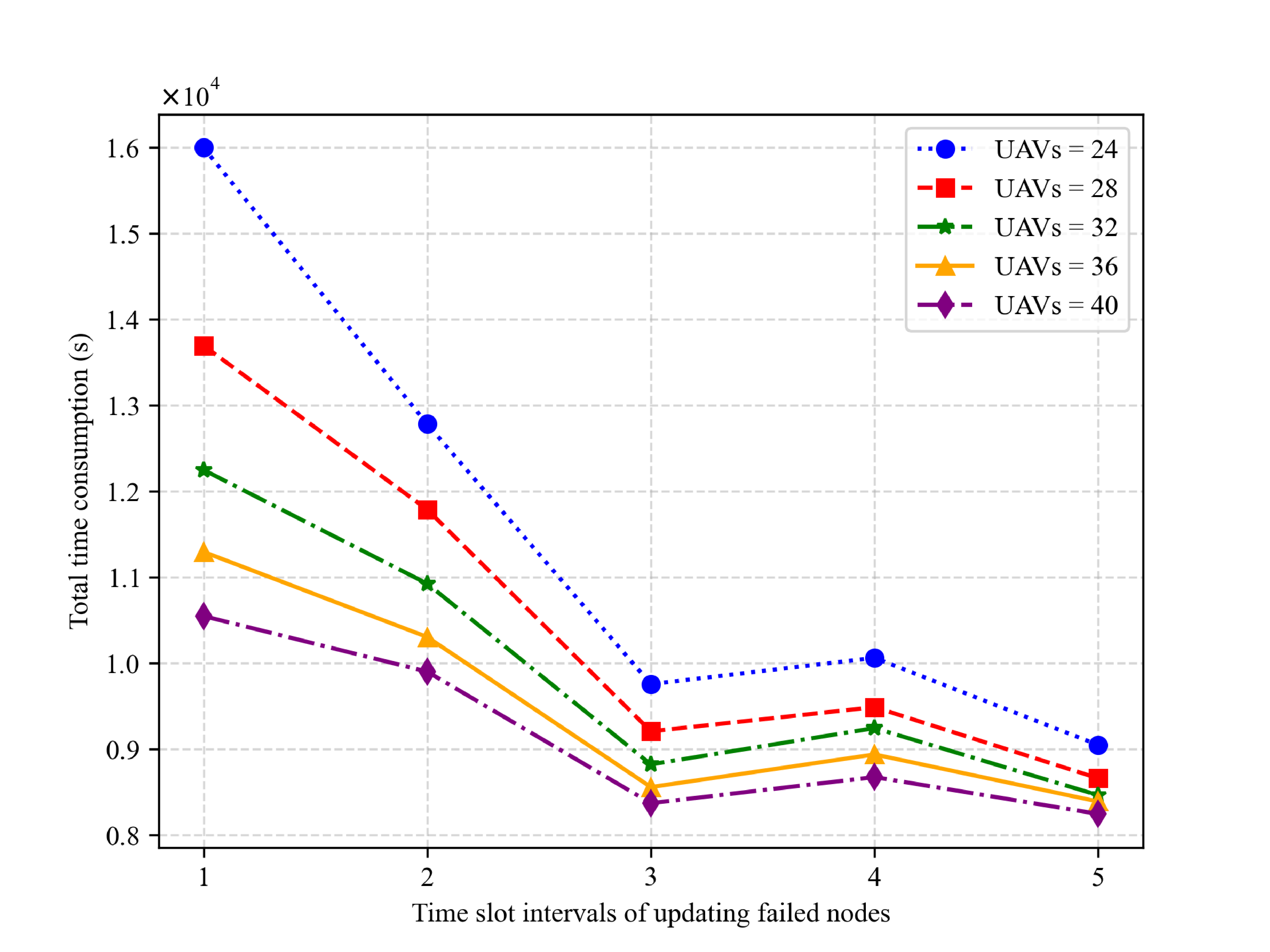}}
	\caption{Total time consumption with different time slot intervals of updating failed nodes.}
    \label{fig:timeslotchange}
\end{figure}

As shown in Fig. \ref{fig:uavchange}, when the number of SFC is constant, the average time consumption per SFC gradually decreases as the number of UAVs increases, because the number of available nodes also grows, and the number of affected SFCs decreases. Moreover, as the number of SFCs increases, the average time consumed per SFC increases, due to the growth of the number of SFCs affected by failed nodes.

Then, we investigate the maximum completion time of an SFC under different number of failed nodes, which is generated according to Poisson distribution in each time slot. Moreover, the maximum completion time denotes the time period from the time the first SFC starts transmitting to the time when all SFC are deployed. As shown in Fig. \ref{fig:nodechange_timeslot}, when the number of UAVs is unchanged and only the parameter of Poisson distribution $\lambda$ increases, the maximum completion time for a single SFC gradually increases. It is accounted that with the increment of $\lambda$, the number of failed nodes increases, and an SFC is more vulnerable to multiple resource failures, thus affecting the maximum completion time. In addition, when the number of UAVs is changed and $\lambda$ is fixed, i.e., the invariable distribution probability of the number of failed nodes, it is noted that the time consumption decreases with the increment of the number of UAVs. Moreover, when $\lambda$ is large, the time consumption decreases more with the growing number of UAVs. This is because when the number of UAVs increases, the number of available nodes that SFCs can choose increases, correspondingly reducing the waiting time of SFCs.

Fig. \ref{fig:dataamount} shows the comparison for different data amounts of SFCs. It is observed from Fig. \ref{fig:dataamount}(a) that when the number of SFCs is unchanged, and the data amount is changed, the sum of time consumption for all SFCs is different. When the data amount increases, the time consumption grows, which has a similar downward trend with the increment of UAVs.

Then, we consider the ratio of SFCs being processed to all unfinished SFCs per time slot, i.e., the number of SFCs being processed in each time slot as a proportion of all unfinished SFCs, which is related to the total time consumption. If the ratio is low, it indicates that a lot of meaningless time consumption is generated, such as waiting, retransmission, and redeployment, rather than prioritizing processing. Hence, we use this ratio to analyze the process of SFC transmission and processing. As shown in Fig. \ref{fig:dataamount} (b), the increment of the number of UAVs can steadily improve the proportion. It is noted that with the growth of the number of available nodes, the number of optional nodes increases, and the number of SFCs affected by failed nodes reduces. In addition, the proportion grows with the increment of data amount, because SFCs needs to be processed for a longer time. Therefore, for each time slot, the number of SFCs being processed increases correspondingly.

\begin{figure*}[!t]
	\centering
	\subfloat[]{\includegraphics[width=.95\columnwidth]{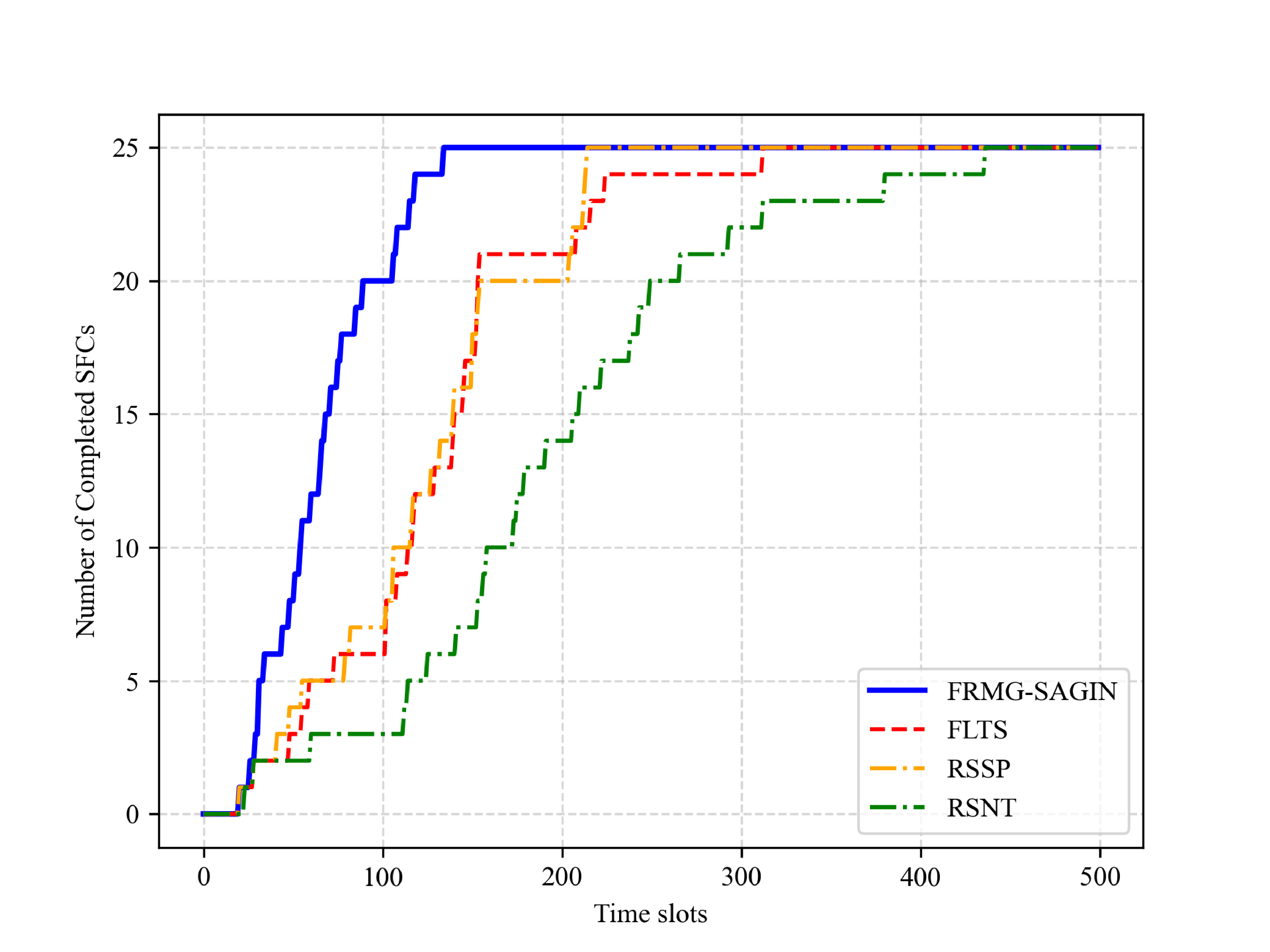}} \label{fig:timeslot}\hspace{7mm}
    \subfloat[]{\includegraphics[width=.95\columnwidth]{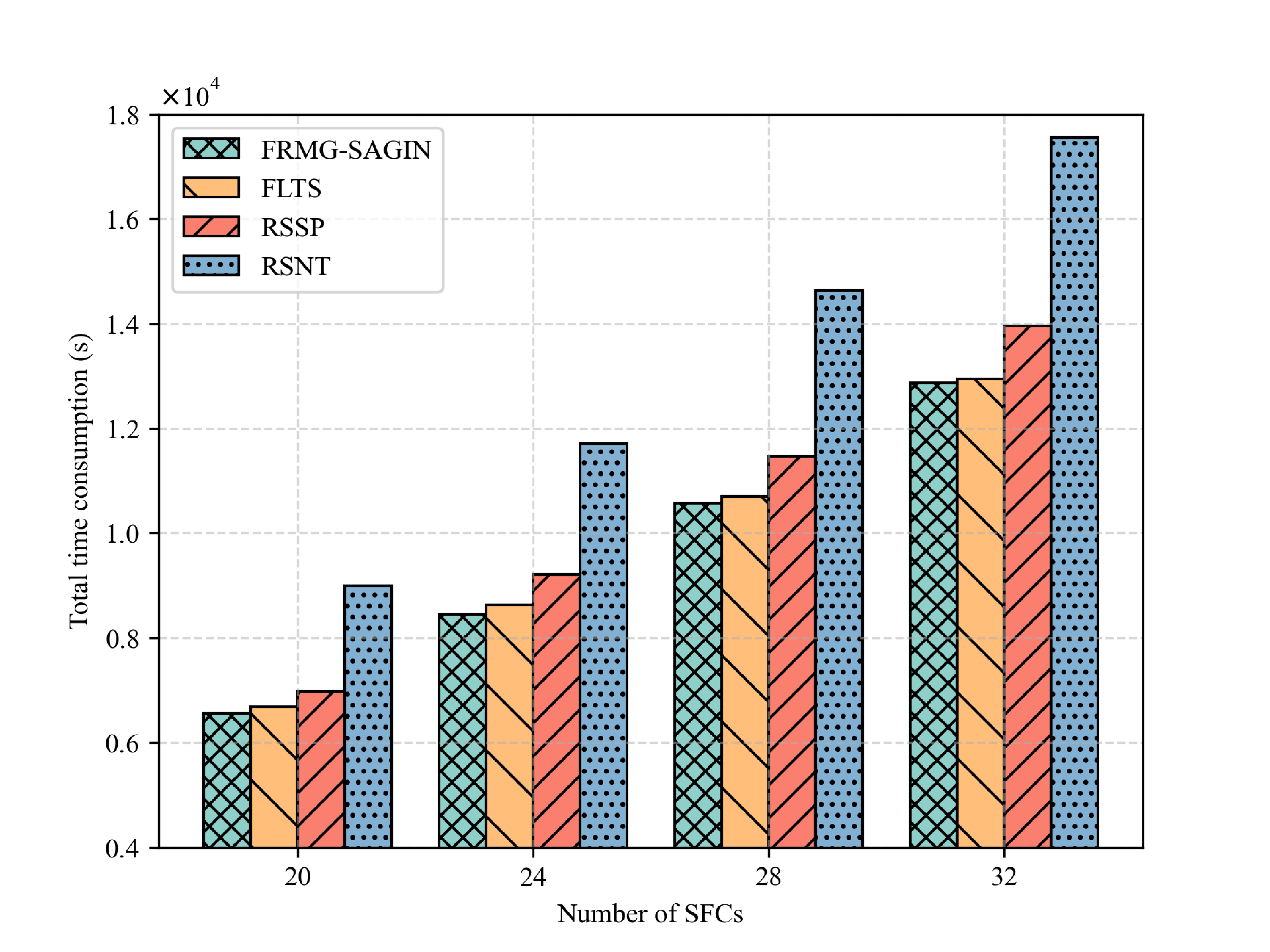}} \label{fig:comparenumber}\\
    \subfloat[]{\includegraphics[width=.95\columnwidth]{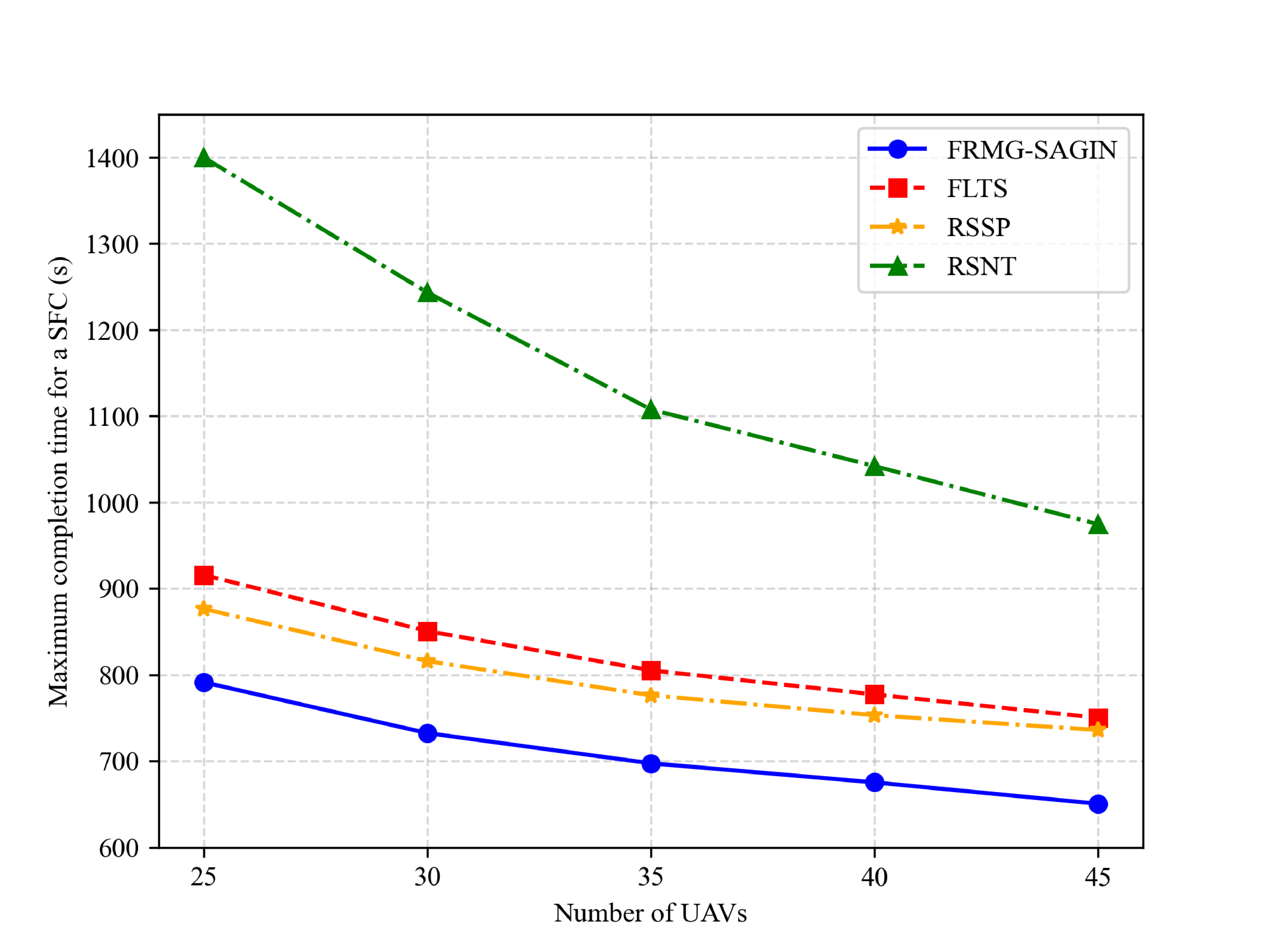}} \label{fig:comparetimeslot}\hspace{7mm}
    \subfloat[]{\includegraphics[width=.95\columnwidth]{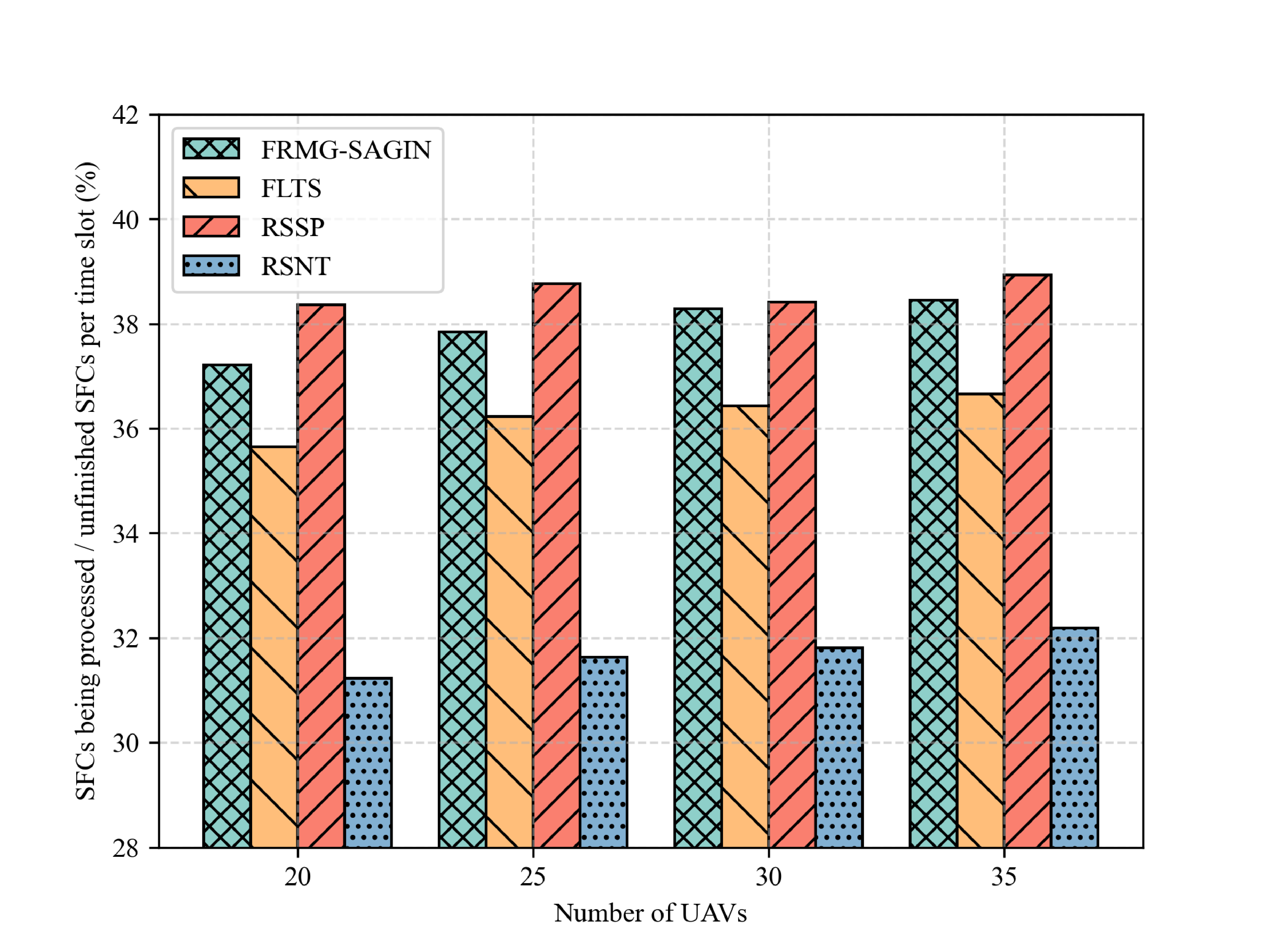}} \label{fig:compareratio}
	\caption{Comparison of various algorithms. (a) The number of SFCs completed deployments with the increasing time slots. (b) Total time consumption with different SFC numbers. (c) Maximum completion time for an SFC with different UAV numbers. (d) Ratio of SFCs being processed to unfinished SFCs per time slot under different UAV numbers.}
    \label{fig:compare}
\end{figure*}

As shown in Fig. \ref{fig:vnfchange}, the maximum completion time varies with different ranges of VNF numbers in each SFC. The horizontal coordinate denotes the range of the number of VNFs in an SFC, i.e., at each loop, the number of VNFs is randomly generated in this range. With the number of VNFs growing, the maximum completion time increases steadily. Moreover, the increment of the number of UAVs reduce the maximum completion time.

Then, we discuss the effect of the updating interval of failed nodes on the total time consumption of SFCs in Fig. \ref{fig:timeslotchange}. When the update interval is relatively short, the impact on the sum of time consumption is large. When the number of UAVs increases, the sum of time consumption decreases rapidly. However, when the situation of node failure is set to be updated every 5 time slots, the sum of time consumption changes slightly with the growing number of UAVs. This is consistent with the performance of network stability.

\subsubsection{Comparison of different algorithms}
We evaluate three different algorithms with the proposed FRMG-SAGIN in Fig. \ref{fig:compare}. The first algorithm involves that in the preferred list of nodes, unlike (\ref{nodeprelist}) in FRMG-SAGIN, SFCs are sorted according to the data amount from largest to smallest (FLTS). The second algorithm is that nodes randomly select SFCs to process (RSSP) when multiple SFCs arrive simultaneously. Another algorithm is that when an SFC is impacted by node failures, it randomly selects the next node for transmission (RSNT). Specifically, the difference between RSSP and RSNT is that nodes random select SFCs in RSSP, and SFCs randomly choose nodes in RSNT.

As shown in Fig. \ref{fig:compare}(a), when all variables and the condition of failed nodes are the same, the algorithms have different deployment processes for a specific number of SFCs. The proposed FRMG-SAGIN takes the shortest time to complete all SFC deployments, while RSNT takes the longest time. The time consumed by the other two algorithms FLTS and RSSP is between the time consumption of FRMG-SAGIN and RSNT. It is observed that RSNT goes through multiple long time periods in which the value of the ordinate is not changed, i.e., no new SFC is completed, which significantly affects the overall time consumption.

As shown in Fig. \ref{fig:compare}(b), the total time consumption when all SFC are deployed is compared, with the least time consumed in FRMG-SAGIN and the most consumed in RSNT. With the increment of  the SFC numbers, the total time consumption increases. The time consumption of FRMG-SAGIN, FLTS and RSSP increases less, while the time of RNST increases more.

In Fig. \ref{fig:compare}(c), the maximum completion time for an SFC is compared. FRMG-SAGIN consumes the shortest time, while RSNT consumes the longest, and larger than other algorithms. Moreover, when the number of UAVs is small, the result of RSNT is the worst, i.e., the algorithm has poor adaptability in small-scale networks, while FRMG-SAGIN can be adapted to networks of all sizes.

Fig. \ref{fig:compare}(d) compares the ratio of SFCs being processed to all unfinished SFCs per time slot in four algorithms. RSSP has the highest utilization ratio, and RSNT has the lowest, which is much lower than the other three algorithms. This is because in RSSP, nodes randomly select the arriving SFCs, and do not reject any SFC. Hence, the SFCs being processed in each time slot accounts for a relatively large proportion. However, when it is analyzed together with Figs. \ref{fig:compare}(b) and (c), it is noted that such random selection of SFCs does not help reduce the overall time consumption. Therefore, FRMG-SAGIN perfects best  in terms of both time consumption and node utilization.

\section{Conclusions\label{sec:Conclusions}}
In this paper, we presented the RTEG framework to address the highly dynamic environment and complex resource allocation challenges in SAGIN. We established a model of SFC deployment and recovery to deal with resource failures. The SFC deployment issue was formulated with the objective of minimizing the total time required to complete all SFC deployments. To tackle the robust recovery problem, we proposed an algorithm based on the many-to-one two-sided matching game. Simulation results confirmed the effectiveness and benefits of the proposed algorithm, demonstrating the time consumption reduction of about 25\% compared to other benchmark algorithms. In the future, we will consider updated solutions and algorithms that are more suitable for the ultra-large scale of SAGIN, to balance the computational cost and performance.

\textcolor{black}
{\bibliographystyle{IEEEtran}
\bibliography{ref}
}

\end{document}